\documentclass[british,table,graybox]{svmult} %% table is there for \rowcolors
\usepackage[utf8]{inputenc}

\usepackage{type1cm}          % SPRINGER
\usepackage{newtxtext}        % SPRINGER
\usepackage[varvw]{newtxmath} % SPRINGER
\usepackage{makeidx}          % SPRINGER
\usepackage[bottom]{footmisc} % SPRINGER
\usepackage[british]{babel}
\usepackage{csquotes}         % quotes
\usepackage{xparse}           % WHAT FOR ?
\usepackage{xspace}
\usepackage{xcolor}
\usepackage{graphicx}         % SPRINGER
\graphicspath{{figures/}}     % figures folder
\usepackage{tikz}
\usepackage{array}
\usepackage{booktabs}
\usepackage{tabularx}
\usepackage{colortbl} 
\definecolor{lightgray}{gray}{0.93}
\usepackage{multicol}         % SPRINGER
\usepackage{pdflscape}        % FOR THE FAILURE SUMMARY
\usepackage[hidelinks,bookmarks=false,breaklinks=true]{hyperref}
\usepackage[capitalise,nameinlink,noabbrev]{cleveref}
\usepackage{xurl}
\usepackage[sort&compress,numbers]{natbib}
\usepackage{bibtopic}        % split bibliography
\usepackage{algpseudocode}   %% FOR PSEUDO CODE OF HAYMAN ALGORITHM
\usepackage{enumerate}
\usepackage{paralist}

\usepackage[textsize=scriptsize,textwidth=4.3cm]{todonotes} %% for \todo
\definecolorseries{test}{rgb}{grad}[rgb]{.95,.55,.55}{11,11,17}
\resetcolorseries[10]{test}
\newcommand{\addtodoeditor}[1]{%
    \colorlet{#1}{test!!+!50}
    \expandafter\newcommand\csname#1\endcsname [1]{%
        \todo[color=#1,size=\tiny]{\sffamily\textbf{{#1}:}
    ##1}\xspace%
    }
    \expandafter\newcommand\csname#1i\endcsname [1]{%
        \todo[inline, color=#1]{\sffamily\textbf{{#1}:} ##1}\xspace%
    }
}
\addtodoeditor{af}
\addtodoeditor{gb}
\addtodoeditor{mpr}
\addtodoeditor{mrm}

\usepackage[frozencache=true,cachedir=_minted]{minted}
\setminted{fontsize=\normalsize}

\usepackage[en-GB]{datetime2}
\RenewDocumentCommand{\date}{o m m}{%
  \IfNoValueTF{#1}{%
    {\DTMenglishmonthname{#2}, #3}%
  }{%
    {\DTMdate{#3-#2-#1}}%
  }%
}

\newcommand{\USD}{USD\xspace}
\newcommand{\GBP}{GBP\xspace}
\newcommand{\EUR}{EUR\xspace}

\newcommand{\trillion}{trillion\xspace}
\newcommand{\trillionS}{\underline{trillions}\xspace} %% plural
\newcommand{\tns}{\underline{tn}\xspace} %% singular
\newcommand{\billion}{billion\xspace}
\newcommand{\billions}{billion\xspace} %% singluar
\newcommand{\billionS}{billions\xspace} %% plural
\newcommand{\bns}{bn\xspace} %% singular
\newcommand{\million}{million\xspace}
\newcommand{\millions}{million\xspace} %% singular
\newcommand{\millionS}{million\xspace} %% plural
\newcommand{\money}[3]{#1~\ensuremath{#2}~#3} %% CURRENCY AMOUNT SCALE (this macro
\newcommand{\church}{\href{https://en.wikipedia.org/wiki/Alonzo_Church}{Alonzo Church}\xspace}
\newcommand{\turing}{\href{https://en.wikipedia.org/wiki/Alan_Turing}{Alan Mathison Turing}\xspace}
\newcommand{\ocaml}{\href{https://ocaml.org}{Ocaml}\xspace}
\newcommand{\haskell}{\href{https://www.haskell.org/}{Haskell}\xspace}

\newcommand{\formalmethods}{formal methods\xspace}
\newcommand{\Formalmethods}{Formal methods\xspace}
\newcommand{\FormalMethods}{Formal Methods\xspace}

\newcommand{\formalverification}{formal verification\xspace}
\newcommand{\Formalverification}{Formal verification\xspace}
\newcommand{\FormalVerification}{Formal Verification\xspace}

\usepackage{orcidlink}
\def\orcidID#1{\unskip$^{\orcidlink{#1}}$}

\begin{document}

\newtheorem{exercice}{Exercise}

%\title*{The costs of faulty software systems}
\title*{Software is Infrastructure: Failures, Successes, Costs, and the Case for \FormalVerification}

\author{
  Giovanni Bernardi\orcidID{0009-0008-3653-3040} \and
  Adrian Francalanza\orcidID{0000-0003-3829-7391} \and 
  Marco Peressotti\orcidID{0000-0002-0243-0480} \and
  Mohammad Reza Mousavi\orcidID{0000-0002-4869-6794}
}
\authorrunning{G.~Bernardi, et al.}
\institute{
  Giovanni Bernardi \at Université Paris Cité, France, 
  \email{gio@irif.fr}
  \and 
  Adrian Francalanza \at University of Malta, Malta,
  \email{adrian.francalanza@um.edu.mt}
  \and 
  Marco Peressotti \at University of Southern Denmark, Denmark,
  \email{peressotti@imada.sdu.dk}
  \and
  Mohammad Reza Mousavi \at King's College London, United Kingdom,
  \email{mohammad.mousavi@kcl.ac.uk}
}

\maketitle

\abstract{
  In this chapter we outline the role that software has in modern
  society, along with the staggering costs of poor software quality.
  To lay this bare, we recall the costs of some of the major software
  failures that happened during the last~$40$ years.
  We argue that these costs justify researching, studying and applying
  formal software verification and in particular program analysis.
  This position is supported by successful industrial experiences.
}

\renewcommand{\paragraph}[1]{\vspace{1em}\phantomsection\noindent{\normalfont\bfseries #1.\xspace}}
\newcommand{\parcase}[2]{\paragraph{#1}\label{#2}}

\label{ch:zero} %% LABEL REQUESTED BY EDITORS
\section{Introduction}
\label{sec:introduction}

%%% SOFTWARE HAS A STAGGERING IMPACT ON NATIONAL ECONOMIES (OUTLINE)
``%Our civilisation runs on software [\ldots]
Software is everywhere, not just in computers but in household
appliances, cars, aeroplanes, lifts, telephones, toys and countless
other pieces of machinery.''
\href{https://www.economist.com/}{The Economist} opened thus an
article published in 2003~\cite{bugtrap}.

Fast-forward to $2011$, and 
\href{https://en.wikipedia.org/wiki/Marc_Andreessen}{M. Andreessen} in the 
\href{https://www.wsj.com/}{Wall Street Journal}
argues that ``software is eating the world'' \cite{andreessen2011},
then in $2014$ Bill Curtis observes in \cite{curtis2014} that
software applications have entered the
era of 9-digit (\money{\USD}{\geq 100}{\million}) software failures, and that
``Not only are these huge figures, but they represent money that is—in
a very real way—purely~wasted.''
Fast-forward again, and in $2020$ \href{https://csacademy.mst.edu/people/members/herbkrasner/}{Herb Krasner} observed that
the cost of finding and fixing bugs
or defects is the largest single expense element in the software
lifecycle \cite{cisq2020}: over a 25-year life expectancy of a large
software system, almost fifty cents out of every dollar will go to
finding and fixing bugs \cite{cisq2020}.

In $2022$ the cost of software failures in the US alone was estimated to
amount to a staggering \money{\USD}{1.56}{\trillion}  \cite{cisq2022}.
%%% NEW SENTENCE
These costs cry out for the application of software validation
techniques to identify existing defects and avoid introducing new ones.
One such technique is software verification via \formalmethods.

%%% OLD SENTENCE
%% These staggering costs cry out for the application of
%% software verification techniques to  identify existing software defects and avoid introducing new ones.

%%% FROM SOFTWARE VERIFICATION TO \FORMALMETHODS / MATHEMATHICS
\paragraph{Software verification} %Program verification}
Software comprise programs, their documentation, and the associated
data, like configuration files \cite[Section 1.1]{DBLP:books/lib/Sommerville9}.
The bulk of software verification pertains to programs,
and this is our focus.

Programs are text: humans {\em write} them and {\em read} them.
Buildings, elevators, bridges, cars, air-planes, \ldots instead
are physical objects: humans {\em build} them.
The words {\em soft}ware and {\em hard}ware suggest this difference.
A consequence is that the validation of programs differs
from the validation of other engineering artefacts.

Concrete artefacts are validated via the laws of physics.
A wall, once {\em built}, is {\em de facto} ran by physics,
and it is sufficient to test that it withstands specific
forces to ensure it operates as intended~\cite[Chapter
  11]{structuralconcrete2019}. 
In fact, at design time, i.e. prior to building the artefact,
these forces are usually modelled and accounted for, so the
final artefact usually withstands them (i.e. is correct) by design.

Programs, once {\em written}, require being run by a computer.
Running a program, though, may be impossible prior to deployment,
and validation, more often than not, is based on tests,
which are also software, i.e. they can contain bugs.
Moreover tests may not find bugs. A recent case is a bug
found in
\href{https://en.wikipedia.org/wiki/Amazon\_DynamoDB}{DynamoDB} that
``passed unnoticed through extensive design reviews, code reviews, and
testing'' \cite{Newcombe2015}.
The limitations of testing are pointed out by private and public entities alike:
practitioners at Amazon have remarked in \cite{Newcombe2015} that
``\Formalmethods find bugs in system designs that cannot be found through
any other technique we know of'', and a report by the White House
\cite{whitehousepaper} states that ``testing is a necessary but
insufficient step in the development process to fully reduce
vulnerabilities at scale.''

At times it is necessary to understand the behaviour of a program,
i.e. what it will do when ran, by only reading it.
\begin{example}
Consider the BASH command line \mintinline{bash}{'rm -rf ~'}.
We advise the reader against running this command to understand
what it does. Instead, we suggest the reader to study the
documentation of the program \mintinline{bash}{'rm'} \cite{rmmanpage}.
A similar argument holds for the command line
\mintinline{bash}{'sudo dd if=/dev/zero of=/dev/sd* bs=1M'}.
\end{example}

The necessity to understand what programs do before running them
motivates the research in {\em program analysis}.
Roughly speaking, program analysis lets us prove properties
of program behaviours by associating programs (i.e. texts)
to mathematically defined semantics, and by using deductive reasoning on these semantics.

%%% SOFTWARE AS ARTIFACT, WHICH SCIENCE BEHIND IT?
\begingroup
\begin{table}[t]
%  \hrulefill
\begin{center}  
    % \caption{The comparison proposed by \href{https://www.di.ens.fr/~rival/}{X. Rival} and \href{https://kwangkeunyi.snu.ac.kr/}{K. Yi} in their
  %   book \cite{rival:hal-02402597}.}
    \caption{Comparison of computing and other engineering areas proposed \citeauthor{rival:hal-02402597} \cite[Fig.~1.1]{rival:hal-02402597}.}
    \label{table:rival-yi}
    \begin{tabularx}{\linewidth}{|l||>{\raggedright\arraybackslash}X|>{\raggedright\arraybackslash}X|}
      \hline
          & Computing areas & Other engineering areas
      \\ \hline \hline
      Object & Software & Machine, building, circuit, chemical process
      \\ \hline
      Execution subject & Computer runs it & Nature runs it
      \\ \hline
      Our question & Will it work as intended? & Will it work as intended?
      \\ \hline
      Our knowledge & Program analysis & Newtonian mechanics, Maxwell equations, Navier-Stokes equations, thermodynamic equations, and other principles
      \\ \hline
    \end{tabularx}
\end{center}
    \hrulefill
\end{table}
\endgroup

An analogy between the creation of software artefacts
and architectural artefacts has been hinted at in
\cite{10.1145/2736348}, while a comparison between
program analysis and the analysis of the artefacts
produced in other engineering fields has been proposed
by \href{https://www.di.ens.fr/~rival/}{X. Rival} and \href{https://kwangkeunyi.snu.ac.kr/}{K. Yi} in \cite{rival:hal-02402597} via
the \cref{table:rival-yi}, which suggests that program analysis should
be treated similarly to the mathematical theories that underpin other
engineering fields.

%%%%%%%%%%%%%%%%%%%%%%%%% FUNCTIONAL CODE
The following example presents archetypal questions one may want to answer {\em before}
running a program.
  
\begin{example}
  \label{ex:factorial}
  Consider the following code snippets which are two {\em tentative} implementations of the factorial function in \ocaml (left) and \haskell (right),\\[-5pt]
  {\centering%
\begin{tabular}{p{0.5\textwidth}p{0.5\textwidth}}
\begin{minted}[frame=single]{ocaml}
let rec fact n =
   if n = 0
   then 1
   else n * (fact (n-1))
  \end{minted}
&
\begin{minted}[frame=single]{haskell}
fact n =
   if n == 0
   then 1
   else n * (fact (n-1))
  \end{minted}
\end{tabular}
  }
The syntactic difference between the two codes,
i.e.~\mintinline{ocaml}{=} in \ocaml and~\mintinline{haskell}{==} in \haskell, 
is not relevant for our discussion, because these two syntaxes
have the same meaning, i.e. logical equality.

According the the Ocaml interpreter
\begin{minted}[frame=single]{ocaml}
  # fact 30;;
  - : int = 458793068007522304
\end{minted}
and according to the Haskell interpreter
\begin{minted}[frame=single]{haskell}
  Prelude> fact 30
  265252859812191058636308480000000
\end{minted}
The computed results are remarkably different.
Which one of the two programs, if any, implements
the factorial function? Why? For which reason?
\end{example}

Program analysis help us answer such
questions as the ones in \cref{ex:factorial}.
In particular, the last question in that example
essentially asks for {\em a proof}, and so mathematical
reasoning is necessary to answer it.
In particular a  mathematically defined semantics of
programming languages and deductive reasoning on it
are necessary.

\paragraph{\Formalmethods} % 
Unambiguous reasoning techniques based on rigorous mathematical models
of the phenomenon under study, in our setting software correctness, are
what we call {\em \formalmethods}.
They are virtually common to every scientific field.

{\em \Formalverification} is software validation carried out via \formalmethods.
Discussing \formalverification without discussing \formalmethods can thus hardly be done.
For instance the appendices of both books \cite{rival:hal-02402597,10.5555/555142}
are devoted to the mathematics that suffice to develop the verification
techniques explained in those books.
The tight connection between \formalverification and \formalmethods
is the reason why the discussion in this chapter oscillates between
the two topics, i.e. we use ``\formalmethods'' and ``\formalverification''
somewhat interchangeably.
Further clarification of the terminology can be found in \cref{sec:glossary}.

%Aside from being necessary to answer certain questions,
%mathematical reasoninghas
\Formalmethods have two advantages. 
First, they help %it helps
building confidence that programs operates as intended,
{\em prior} to deploying them. Mathematical reasoning allows
even {\em proving} that (at least models of) a program behaves as expected.
Tools grounded in mathematical reasoning such as
\href{https://www.key-project.org/}{KeY}  %\cite{KeYtool}
and interactive theorem provers (\href{https://rocq-prover.org/}{Rocq}, %
\href{https://lean-lang.org/}{Lean}, 
\href{https://isabelle.in.tum.de/}{Isabelle/HOL}, etc. \ldots)
have proven to be, now and again, very successful when not {\em
  tout-court} essential.
For instance, according to \cite{brooker2025} %\cite{10.1145/3729175}
``[\ldots] having the ability to formally check (and, in some cases,
prove) that proposed design optimizations are correct allows naturally
conservative distributed-systems engineers to be bolder in their
protocol design choices without increasing risk [\ldots]''.
A similar example is the compiler CompCert \cite{compcertwebpage},
which allowed \href{https://www.airbus.com/en}{AirBus} to improve
the worst-case execution time of (some of) its software by 12\% \cite[Slide 7]{airbusWCET}.
Another success story is the bug found in the sorting algorithms of
Python and Java by the KeY team \cite{DBLP:conf/cav/GouwRBBH15}.

%% Using the HOL Light interactive theorem prover, the team was able to
%% prove the correctness of these optimizations. Given the high
%% percentage of cloud CPU cycles spent on cryptography, this type of
%% optimization can significantly reduce infrastructure costs and aid
%% sustainability while at the same time improving customer-visible
%% performance.``

%\item 
%% The second advantage
%% \todo{A: what was the first advanatage?}
%% of
Second, techniques grounded in mathematical/deductive reasoning
help pass down information through the generations because
\begin{inparaenum}[(a)]
\item mathematical techniques that help solving problems become part of standard
  university curricula, thereby providing a common unambiguous language for practitioners; and
\item they can reduce the costs due to ``Technical Debt'', i.e. short-term and ad-hoc choices which lead to  costs on the
  long term.
\end{inparaenum}
%\end{enumerate}

\Formalmethods in software validation are so important that even political
players started advocating for them. Case in point, a report by the
White House \cite{whitehousepaper} states that
``\Formalmethods [\ldots] allow for proving the presence of an {\em
  affirmative} requirement, rather than testing for
the absence of a {\em negative} condition. [\ldots] They offer a way
to eliminate, not just mitigate, entire bug classes. This is a
remarkable opportunity for the technical community to improve the
cybersecurity of the entire digital ecosystem''

\paragraph{Questions}
Mathematical reasoning on software semantics appears to be costly both in terms of time
and {\em skills} required to perform it.
Our experience in nearly a decade of teaching \formalmethods and
static analysis is that the vast majority of students do not
understand {\em why} a mathematical approach is worth any attention,
and some of them are just not interested in acquiring the skills to
{\em prove} a program correct.
The question that some students seem too afraid to ask
explicitly is 
\begin{equation}
  \label{question1}
  \tag{Q1}
  \text{Why should we spend any time studying \formalmethods?}
\end{equation}
The question is less naïve than it may seem. To see why, %
let us rephrase it in economical %rather than temporal
terms
\begin{equation}
  \label{question}
  \tag{Q2}
  \text{Why should taxpayers money fund research in \formalmethods?}
\end{equation}
%% Even more concretely, we ask the reader, assuming she is a taxpayer,
%% \begin{center}
%%   How much of your taxes would you like to allocate for research on
%%   \formalmethods ?
%% \end{center}

%%% CHAPTER STRUCTURE

This chapter tries to answer questions~(\ref{question1},\ref{question})
by {\em reminding} the reader via publicly available empirical data of
\begin{inparaenum}[(a)]
\item the key role of software systems in our society,
\item the staggering cost of their failures, and
\item the documented return of investment from \formalverification in IT.
\end{inparaenum}

First, in \cref{sec:infrastructure} we argue that software is
so essential to operate the infrastructure of western society,
that it should be considered {\em itself} as infrastructure.
Developing techniques to understand why software works correctly is no less important that
developing techniques to understand why aqueducts and bridges work.
This is our first answer to questions~(\ref{question1},\ref{question}).

The importance of understanding code brings about one obvious
question: how easy a task is it?
In \cref{sec:program-analysis} we present a series of real
programs, and ask the reader to understand them.
Our aim is to outline concretely
how complicated understanding code and even pseudo-code
actually is. There is no short-cut in formal reasoning.
%A person willing to reason on software must be ready to
%invest the required brain-power.

In \cref{sec:costs} we summarise the costs of and the root causes
behind some of the major failures in software (or its industrial operations)
that happened during the last~$40$ years.
The figures we recall in \cref{sec:costs} let us answer 
questions~(\ref{question1},\ref{question}) thus: the
only way to avoid the costs of software failures is prevention.
This ultimately boils down to more careful software validation/\formalverification.
%% In hindsight, investments in the \formalverification of
%% software, and in the fundamental research necessary to
%% conceive and develop suitable verification techniques
%% may have lead to considerable savings on the long term.
%% We hope this evidence will provide one answer to question
%% (\ref{question}).

This remark spurs a natural question: how much should be invested
in validation/\formalverification prior to software deployment?
In \cref{sec:potential-costs} we address this question by outlining
some of the {\em potential} average costs of software failures such
as outages and data breaches.
%due to
%erroneous usage of APIs and of communication protocols.

While measuring the costs of failures is possible thanks to the
ensuing investigations and/or the accounting consequences, 
measuring the merits of \formalverification is more complicated:
on the one hand potential costs may not be convincing, and on
the other hand when bugs are found, they are corrected,
failures are avoided, and thus there hardly is any measurable consequence.
To tackle this difficulty, \cref{sec:mitigation} presents a
series of industrial success stories, along with figures that
support the investment of resources into \formalverification.
We hope the data given in \cref{sec:advantages} will provide one
more answer to questions~(\ref{question1},\ref{question}).

\cref{sec:beyond} reports a positive trait of \formalmethods
that has emerged in education, and argues that the same trait
may help in tackling technical debt.

%%% related works
Given the vastness of the subject matter
(from speeches of presidents to the {\tt NULL} value), our exposition
touches but the surface of a number of different subjects,
that different communities have attentively investigated.
\cref{sec:related-work} thus presents pointers to further material
that the reader may find useful either for historical reasons or to
investigate further the subjects discussed in this chapter.

\cref{sec:future-work} lists a series of avenues for future studies, and 
\cref{sec:conclusion} closes the chapter with a few final considerations.

%% \begin{figure}[t]
%%   \hrulefill
%%   \begin{center}%
%%     \begin{tabular}{ccc}
%%       \href{https://www.youtube.com/watch?v=pDBnQPGc4mQ&t=104s}{\includegraphics[width=100pt]{210412-joe-biden-silicon-wafer.jpg}}
%%       &
%%       \href{https://www.elysee.fr/emmanuel-macron/2025/03/05/adresse-aux-francais-6}{\includegraphics[width=100pt]{allocution_macron.jpg}}
%%       &
%%       \href{https://www.youtube.com/watch?v=g7ufeyfL1u4}{\includegraphics[width=100pt,height=66.5pt]{sunak.jpg}}
%%       \\[1pt]
%%       (a)
%%       &
%%       (b)
%%      &
%%      (c)
%%     \end{tabular}
%%   \end{center}
%%   \caption{%
%%     Column (a) shows %
%%     the $46^{\mathit{th}}$ president of the
%%     United State of America, stating that chips are infrastructure
%%     \cite{BidenThisisinfrastructure}. %
%%     Column (b) shows %
%%     the $25^{\mathit{th}}$ president of the French Republic during the address in
%%     which he cited cyber attacks to hospitals \cite{AllocationMacron05032025}.
%%     Column (c) shows %
%%     the $57^{\mathit{th}}$ prime minister of the United Kingdom giving the address
%%     on the damages and compensations for the Horizon IT Scandal \cite{sunak2024}.
%%   }
%%   \label{fig:thisisinfrastructure}
%%   \label{fig:macron}
%%   \label{fig:sunak}
%%   \hrulefill
%% \end{figure}

\section{Empirical data}
\label{sec:infrastructure}
%\todo{A: Should this be a section instead of a subsection?}

%%% MEANING OF INFRASTRUCTURE
According the Oxford Dictionary of English the word ``infrastructure'' means
"the basic physical and organisational structures and facilities
(e.g. building, roads, power supplies) needed for the operation of a
society or enterprise''.

Software is essential to the operation of western society,
and we argue that, albeit not physical, {\em software is
  infrastructure}.
We support this point via qualitative data that we gather
in \cref{sec:infra-data}, and via aggregated
quantitative data that we  present in \cref{sec:aggregated-costs}.

%%%%%%%%%%%%%%%%%%%%%%%%%%%%%%%%%%%%%%%%%%%%%%%%%%%%%%%%%%%%%%%%%%%%
%%% SOFTWARE IS INFRASTRUCTURE

%\todo{A: perhaps give an outline at this stage by mentioning Data centers, Healthcare, Chips etc}
\subsection{Existing infrastructures}
\label{sec:infra-data}
In this subsection we recall statements by country leaders and
we gather public data on chips, datacenters, and healthcare.

%% CHIPS ACT
\paragraph{Chips}
On \date[12]{4}{2021} the president of the United States of America
stated to the media ``This is infrastructure'' \cite{BidenThisisinfrastructure}
while holding a silicon wafer, a few seconds short of signing the chip act
\cite{wkUSchipact}.
On \date[23]{9}{2023} also the European Union introduced a chip act
\cite{EUchipactpressrelease,EUchipact}.
On the one hand chips like CPU and GPU are engineered in order to
run software; on the other hand languages like
\href{https://fr.wikipedia.org/wiki/Verilog}{\textsf{VERILOG}} and 
\href{https://en.wikipedia.org/wiki/VHDL}{\textsf{VHDL}} 
are crucial to describe the design of chips.
In fact chips descriptions are not only crucial, but complex enough
to contain bugs, consider for instance \cite{DBLP:journals/corr/abs-2108-10771}.

CPU like the ones made by Intel actually run a microcode, a form of
software, that can even be patched, and virtually every processing
unit depends on some form of firmware, i.e. software.
Already in $2007$ the percentage of silicon area taken up by
firmware in chips manufactured below 100 nm processes was over
30\% and raising monthly \cite{moretti07}.

If chips are infrastructure, what should we think of the software
necessary to design them, and that is run by them?

%\gbi{P-E. D. says: the real bottleneck is the ''fab'': they cost billions to build!}

%% DATACENTERS
\paragraph{Datacenters} 
On \date[12]{9}{2024} %\date{9}{12}{2024}
the UK government designated ``UK Data
Infrastructure'' as ``Critical National Infrastructure'' \cite{ukgovt2024,ukparliament2024}.
The UK Managing Director of \href{https://www.equinix.com/}{Equinix}
reacted to this news remarking that
``The internet, and the digital infrastructure that
underpins it, has rapidly grown to be as fundamental to each one of
our daily lives as water, gas, and electricity, and is now a service
that people and the UK economy can no longer live without.''
Indeed, the Department for Science, Innovation and Technology
introduced the ``Cyber Security and Resilience Bill'' \cite{uksecuritybill2025},
according to which in one year half of the businesses reported some form of cyber
security breach or attack \cite{cyberbreachessuvery2024}.

If data centres are critical national infrastructure, and ``Hostile
cyber activity in the UK has grown more intense, frequent, and
sophisticated, with real world impacts for UK citizens''
\cite{uksecuritybill2025}, 
should we not consider the software that actually makes them operate
also part of the infrastructure?

\paragraph{DNS servers}
The Internet is at the core of modern economy and of much of our life-style.
It relies not only on the physical infrastructure
such as submarine cables \cite{submarinecablemaps},
but also on the software that transmits the data from one
computer to the other.
In the complex architecture of the Internet one key software
component is often ignored: the Domain Name System (DNS).
DNS servers world-wide are in charge of transforming URL such as
\url{root-servers.org} into the IP address of the machine to contact
in order to retrieve the required information, for instance {\tt 193.0.11.23}.

The Internet as users know it depends on the DNS being always
available, and on the planet there are $13$ DNS root
servers whose security is a thoroughly studied matter \cite{Baranowski03}.
IDC reported that between $2019$ and $2020$ the average cost per
attack to DNS servers was nearly \money{\USD}{1}{\million}  \cite{IDC-DNS-report2020}.
DNS servers are a core part of modern digital {\em infrastructure}, 
arguably as fundamental as electricity is to buildings.

Should digital infrastructure be any less important than
the rest of our infrastructure?

\paragraph{Healthcare}
On \date[5]{3}{2025} the president of the French Republic gave a
speech~\cite{AllocationMacron05032025} 
that mentions explicitly cyber attacks to block ``our hospitals''.
Software run in hospitals appear to be faulty enough to let these
attack take place: in \cref{sec:attacks-to-hospitals} we gather
a series of links to newspaper articles about a series of attacks that
English and Canadian hospitals underwent.

Here we recall other software issues related to health systems.
In 1992 ambulances in London underwent a
series of problems that lead to an estimated $20 - 30$ casualties
(\cref{sec:costs} presents a detailed account of this failure).
In 2004 the authors of \cite{Zeeberg2004} observed that spreadsheets
may inadvertently change gene names to non-gene names. The problem
still existed in 2021, when \cite{Abeysooriya2021} identified gene
name errors in $30.9\%$ of articles with supplementary Excel gene lists.
In 2016 popular pieces of software for functional MRI have been
found to have false positive rates up to $70\%$, thus invalidating
$15$ years of brain research
\cite{doi:10.1073/pnas.1602413113,mri1,mri2}.
In 2017 pacemaker code has been found to present thousands of bugs
\cite{pacemakers1,pacemakers2}, and vulnerabilities were found
in an automated medication dispensing system \cite{pyxis}.
In 2018, as reported by the BBC, it was estimated that in the UK
between 135 and 270 women may have had their life shortened
because they did not receive invitations to a final routine breast
cancer screening. According to the then UK Health Secretary 
``50,000 women aged 68-71 had failed to get invitations since 2009.'' 
\cite{breastcancer1}, and ``the problem was due to a 
`computer algorithm failure'". According the BBC the error led to
women in the control group wrongly being offered
screenings up until their 70th birthday, rather than their 71st
\cite{breastcancer2}. The final report on the matter states that
the root cause of the incident was a mismatch between how age was
expressed in national service specifications and how age was defined
by the underlying IT system \cite{phefinalreport}.
In 2021 due to a problem in the routing software at Orange,
the emergency phone numbers in France stopped working for 7 hours.
This failure indirectly caused 5 deaths \cite{wkOrangeBug,FranceInfoOrangeBug,NouvelObsOrangeBug}.

If hospitals are part of our infrastructure,
should we not treat the software that helps running them
also as part of our infrastructure?

%% Given the figures and the arguments penned above,
%% we maintain that, albeit not physical, {\em software is
%%   infrastructure}.

%%%% COSTS

\subsection{Aggregated software cost}
\label{sec:aggregated-costs}
Thus far we used rhetorical questions to suggest that we should
treat software as infrastructure because {\em de facto} it already
makes a number of our infrastructures work. 
There is a second reason to treat software as infrastructure:
costs related to software are comparable, when not bigger, to the
ones of other national infrastructures.
We support this position via the figures given in columns TWIS and CPSQ of Table~\ref{tab:aggregated-costs-TWIS},
which highlight the scale of software related costs: \trillionS of \USD.

We gather in %
Table~\ref{tab:aggregated-costs-GDP},
Table~\ref{tab:aggregated-costs-NHS},
and Table~\ref{tab:aggregated-costs-salaries}
other figures to help compare the  the scale of software related costs
to the costs of other national infrastructures.

\begin{table}[t]
    \caption{``Total worldwide IT spending'' (TWIS) and ``cost of poor quality software'' (CPSQ).}
  \label{tab:aggregated-costs-TWIS}
  \centering%
  \begin{tabular}{|c|c|c|c|c|}
      \hline
      year & TWIS & CPSQ & Unit & sources \\
      \hline
      2018 & 3.7 & 2.1 & \tns \USD & \cite{gartner2018}, \cite{cisq2020} \\
      %% !!! cisq2020 above is the correct reference !!!
      \hline
      2020 & 3.9 & 2.6 & \tns \USD & \cite{gartner2020}, \cite{cisq2020} \\
      \hline
      2022 & 4.0 & 2.41 & \tns \USD & \cite{gartner2022}, \cite{cisq2022} \\
      \hline
    \end{tabular}\\
  \hrulefill
\end{table}

\begin{table}[t]
  \caption{GDP in the year~$2024$ of the countries that the authors
    of this chapter work in \cite{imfdata2025}.}
  \label{tab:aggregated-costs-GDP}
  \centering%
      \begin{tabular}{|l|c|c|c|}
      \hline
      Country &  GDP 2024 & Unit \\
      \hline
      Denmark & 429.46 & \bns \USD \\
      \hline
      UK & 3.64 & \tns \USD  \\
      \hline
      France & 3.16 & \tns \USD  \\
      \hline
      Malta & 24.32 & \bns \USD  \\
      \hline
      \end{tabular}\\
      \hrulefill
\end{table}

\begin{table}[t]
        \caption{Annual budget for the English National Health Service; the data in \cite[Table 1]{kingbudget2018} is converted to USD.}
        \label{tab:aggregated-costs-NHS}
  \centering%
      \begin{tabular}{|c|c|c|}
      \hline
      fiscal year & NHS Budget & Unit \\
      \hline
      % 114.6 bns £
      2018/19 & 154.39  & \bns \USD \\
      \hline
      % 121.8 bns  £
      2019/20 & 164.09 & \bns \USD \\
      \hline
      % 128.2 bns  £
      2020/21 & 172.71 & \bns  \USD \\
      \hline
      % 134.4 bns  £
      2021/22 & 181.06 & \bns \USD  \\
      \hline
      \end{tabular}\\
      \hrulefill
\end{table}

\begin{table}[t]
  \caption{Gross salaries of full university professors in four European countries.}
  \label{tab:aggregated-costs-salaries}
  \centering%
    \begin{tabular}{|l|c|c|c|}
      \hline
      & Top yearly gross salary & & \\
      Country &  of a full professor & Unit & Sources\\
      \hline
      %852.645 DKK
      Denmark & 114.30 & k \EUR & \cite{DMdata2025} \\
      \hline
      %105.000  £
      England & 108.47 & k \EUR & \cite{salaries-professors}\\
      \hline
      France & 73.34 & k \EUR & \cite{salaries-professors} \\
      \hline
      Malta & 62.66 & k \EUR & \cite{Maltadata2025} \\
      \hline
    \end{tabular}\\
    \hrulefill
\end{table}

%% \begin{table}[t]
%%   \caption{Figures to compare the scale of software related costs to
%%     the costs of national infrastructures.}
%%   \label{tab:aggregated-costs}
%%   \begin{center}
%%     \begin{tabular}{cc}
%%     &
%%     \\
%%     (a)
%%     &
%%     (b)
%%     \\[10pt]

%%         &

%%     \\
%%     (c)
%%     &
%%     (d)
%%     \end{tabular}
%%   \end{center}
%% \hrulefill
%% \end{table}

%As for \cref{tab:aggregated-costs-TWIS}, we abbreviate ``Total worldwide IT spending''
%with TWIS and ``cost of poor quality software'' with CPSQ.

The data in column TWIS of \cref{tab:aggregated-costs-TWIS} is taken from
the estimations made by \href{https://www.gartner.com/}{Gartner}.
The data in column CPSQ is taken from the reports by the
\href{https://www.it-cisq.org/}{Consortium for Information \& Software
  Quality} CISQ.
Since the players who produced these data differ, columns TWIS and
CPSQ should {\em not} be compared directly against each other.
For instance, according to the CISQ report \cite{cisq2018} the total
worldwide IT spending in~$2018$ was of~\money{\USD}{6.3}{\trillion},
while according to Gartner \cite{gartner2018} it is of~\money{\USD}{3.7}{\trillion}.

%% For comparison, column~(b) reports the GDP in~$2023$ of the countries that the authors
%% of this chapter work in \cite{imfdata2025}; %
%% column~(c) of \cref{tab:aggregated-costs} reports the annual budget
%% for the English National Health Service (we converted the data in
%% \cite[Table 1]{kingbudget2018} to USD); %
%% and  column~(d) reports the gross salaries of full university
%% professors in four European countries.

\paragraph{The lack of quality is expensive} 
We would like the reader to focus on the appalling figures shown in
column (CPSQ) of \cref{tab:aggregated-costs-TWIS}.
\href{https://www.economist.com/}{The Economist}
in $2003$ \cite{bugtrap} was correct:
``In a society dependent on software, the consequences of programming errors (“bugs”) are becoming increasingly significant.''

The first statistics to support this position were provided
by the American National Institute of Standards \& Technology, which
in \cite{NISTreport2002} estimated that software bugs were so common
that their cost to the American economy alone was \money{\USD}{60}{\billion} a
year or about $0.6$\% of gross US product.
This was in $2002$, i.e. before the invention of smartphones.\footnote{In $2023$ the number of mobile phones surpassed the number of people in the world \cite{numberofphones}.}

Since then the situation got worse. According to \cite{cisq2018} 
\href{https://www.tricentis.com/}{Tricentis} analysed~$606$ software
fails from~$314$ companies to understand the business and
financial impact of software failures. The report revealed
that these software failures affected~$3.6$ \billion people and
caused~\money{\USD}{1.7}{\trillion} in financial losses and a
cumulative total of~$268$ years of downtime.
Software bugs were the most common reason behind these failures.

Finally, \cite{cisq2022}
estimated in $2022$ that  the costs of software failures amounting to
a staggering~\money{\USD}{1.56}{\trillion}, i.e. more than half of the total
estimated CPSQ.

%%%% OLD MATERIAL. DO NOT DELETE BUT IGNORE.
%% \paragraph{How consequential are software induced failures ?}

%% \todo{Write an introductory paragraph about the Sleiper-A.}

%% \paragraph{Replication of failures.}
%% One issue posed by software as artifact, it that thanks to the
%% Internet it can be replicated and deployed at essentially no cost.
%% The CrowdStrike faulty update has deployed in less than 24 hours to
%% $8.5$ computers in $19$ different countries, impacting
%% hospitals and airports (just to name parts of critical
%% infrastructure).

%% By contrast, the CFM
%% LEAP\footnote{\url{https://en.wikipedia.org/wiki/CFM_International_LEAP}}, 
%% a succesful airplane engine,
%% has been launched in July $2008$. Between this year and $2019$
%% thourgh its different iterations (1A, 1B, 1C),
%% it has been produced (i.e. replicated and deployed)
%% $2,516$ time.
%% This number pales in comparison to the speed and the cost with with
%% software can be replicated and deployed.

\section{Challenges for computer science students}
\label{sec:program-analysis}
A program is a text. Since running a program may have catastrophic
consequences, in general we need to be able to {\em understand}
what a program does {\em before} running it, that is just by reading
it.
Indeed, developers spend $70$\% of their time understanding code,
and only $~5$\% editing it \cite{DBLP:conf/iwpc/MinelliML15}.

This section presents minimal programs and the pseudo-code
of an algorithm to challenge the reader: how easy is it to understand
them?
% OLD TEXT
%The material and the questions should be clear to third year
%undergraduate students in CS.
Exercises~\ref{exa:fp} to~\ref{exa:zune} are sequential code,
and should be clear to third year undergraduate students in CS.
Exercice~\ref{exa:hyman} instead pertains to concurrency theory,
and should be accessible to master and graduate students.

As warm up we propose the following example.
\begin{exercice}[Python]
  \label{exa:fp}
Which value does the following code print (if any)?
\begin{minted}[frame=single]{python}
a = 0.0
while (a != 100.0):
    a += 0.1
print(a)
\end{minted}
\end{exercice}

\begin{exercice}[C]
  \label{exa:fast-inverse-sqrt}
  Which mathematical function is the method \mintinline{C}{mystery}
  given below supposed to implement and, more importantly, why?
  \begin{minted}[frame=single]{cpp}
float mystery( float number )
{
  long i;
  float x2, y;
  const float threehalfs = 1.5F;

  x2 = number * 0.5F;
  y  = number;
  i  = * ( long * ) &y;
  i  = 0x5f3759df - ( i >> 1 ); 
  y  = * ( float * ) &i;
  y  = y * ( threehalfs - ( x2 * y * y ) );

  return y;
}
  \end{minted}
\end{exercice}

The code in Exercise~(\ref{exa:fast-inverse-sqrt}) is not obfuscated,
and it has played a key role in the video game industry near the end
of the $90$'s.

%%%%%%%%%%%%%%% FLAKY TESTS
The reader may argue that the code in
Exercise~(\ref{exa:fast-inverse-sqrt}) is hard to understand because it uses pointers, which are a notorious source of headaches,
and because the syntax in it is unwieldy.
Consider then the test in the next exercise. Its code has 
no pointers, thus it is supposedly simpler to understand than the code in
Exercise~(\ref{exa:fast-inverse-sqrt}).

\newcommand{\test}{\mintinline{python}{test_json_serialization}}

\begin{exercice}[Python]
  \label{exa:flaky}
  \mbox{ }
\begin{minted}[frame=single]{python}
import json

def test_json_serialization():
    data = {"name": "Alice", "age": 30,
            "hobbies": ["reading", "cycling"]}
    expected_json = """{"age": 30,
                      "hobbies": ["reading", "cycling"],
                      "name": "Alice"}"""
    assert json.dumps(data) == expected_json
\end{minted}
Owing to the \mintinline{python}{assert}, the program
%\mintinline{python}{test_json_serialization}
\test\ is a test:
it checks that the function
\href{https://docs.python.org/3/library/json.html}{\mintinline{python}{dumps}}
computes a given string.
Consider any sequence of consecutive executions of the test
\test\ in the Python interprepter.
One may think that the \test\ returns the same
result in each one of these executions.
This may not be the case. Why?
\end{exercice}

%% THE ZUNE BUG IDE DESCRIBED HERE: https://en.wikipedia.org/wiki/Zune_30

Another interesting example comes once again from real life.
In 2008 Microsoft tried to market their Zune player as a contender to Apple's
Ipod. On \date[31]{12}{2008}, though, all Zune 30 GB
MP3-Player stopped responding.
The root cause of the problem was a software bug in the conversion routine
from system time to calendar time. We give a model of it in the next example.

\begin{exercice}[Java]
  \label{exa:zune}
  Consider the following code, where the variable \mintinline{Java}{leap},
  \mintinline{Java}{days}, \mintinline{Java}{year} are integers.
  \begin{minted}[frame=single]{Java}
while (days > 365) {
  if (isLeapYear(year)) {
    if (days > 366) {
      leap = leap + 1;
      days -= 366;
      year += 1;
    }
  } else {
    days -= 365;
    year += 1;
  }
}
  \end{minted}
  Which is the bug contained in this code?
  In reasoning the reader can consider the variables non-negative, and the function \mintinline{Java}{isLeapYear} correct,
  i.e. it returns \mintinline{Java}{true} if and only if the value of
  \mintinline{Java}{year} (a) can be divided by $4$ {\em and} (b) it cannot
  be divided by $100$ {\em except} if it is divisible by $400$.
\end{exercice}

%%%%%%%%%%%%%%%%%%%%%%%%% PSUEDO CODE AND CONCURRENCY

The reader may contend that the examples given thus far pertain actual code,
while only {\em models} of real programs can be understood by reading them.
To challenge this position, we present a example from concurrency theory.

\paragraph{Concurrency} %
A fundamental safety property of concurrent software is {\em mutual exclusion}. According to the Turing lecture \cite{10.1145/2771951}, mutual exclusion
amounts to ensuring that no two programs execute concurrently their critical sections.
A number of algorithms to enforce mutual exclusion have been designed, and we now focus on one of them.

\begin{exercice}[Pseudo-code]
  \label{exa:hyman}
  Consider the following pseudo-code. It describes an algorithm meant
  to ensure mutual exclusion of multiple processes from a critical section.
  In this particular case, we assume that  $i,j \in \{1, 2\}$ and that
  $i \neq j$, the pseudo-code pertains indeed to two different processes.
  \begin{center}
  \fbox{
    \begin{minipage}{.95\linewidth}
  \begin{algorithmic}[1]
    \While{\texttt{\bfseries true}}
    \State \text{`noncritical section';}
    
    \State $b_i \gets$ \texttt{\bfseries true};
    \While{$k \neq j$}
    \While{$b_j$}
    \State \texttt{\bfseries skip};
    \EndWhile;
    \State $k \gets i$;
    \EndWhile
    \State \text{`critical section';}
    \State $b_i \gets$ \texttt{\bfseries false};
    \EndWhile
  \end{algorithmic}
  \end{minipage}
    }
    \end{center}
  Which are the mistakes in the pseudo-code?
  In reasoning the reader can assume as initial values $k = 0$,
  $b_i = \texttt{\bfseries false}$.
One possible first step to answer is the (mandatory!) Exercise 7.3
of \cite{Aceto_Ingolfsdottir_Larsen_Srba_2007}.
A detailed analysis can be found in \cite{knuth_2023_pj7sh-jyr86}.
\end{exercice}

Should the reader quickly answer the above questions, we provide
additional examples in \cref{sec:appendix-program-analysis}.

%%%%%%%% RATIONALE
We do not answer the above questions on purpose:
we would like to reader to temporarily put this chapter aside,
and go through the intellectual process necessary to answer them,
possibly keeping track of how much time is spent in the exercise.

Finding the answers to the above questions is in fact already complex
for the code of \mintinline{C}{mystery}, which is but $10$ lines long (including declarations),
and even for the tentative implementations of the factorial function,
that are virtually one-liners.

%%This is how complicated program analysis is.

\begin{table}[t]
  %\hrulefill
  \caption{Estimates of the sizes of some major well-know programs. The lines ``languages$_{> 0.0\%}$'' contain the number of languages used in the software projects at hand, that have more than $0.0\%$ occurrences in the respective codebases.}
  \label{tab:software-size}
  \begin{center}
    \begin{tabular}{|l|c|l|c|}
  \hline
  \multicolumn{2}{|c|}{Linux kernel \cite{LOClinux}} &   \multicolumn{2}{c|}{GNU C library \cite{LOCglibc}}\\
  \hline
  \# LOC & $> 27 \times 10^6$
  &
  \# LOC & $\sim 1.9 \times 10^6$ \\
  \hline
  \# languages$_{> 0.0\%}$ & 8
  &
  \# languages$_{> 0.0\%}$ & 9 \\
  \hline
  \hline
  \multicolumn{2}{|c|}{{\tt Firefox} \cite{LOCfirefox}} &   \multicolumn{2}{c|}{{\tt Blender3D} \cite{LOCblender}}\\
  \hline
  \# LOC & $> 44 \times 10^6$
  &
  \# LOC & $5 \times 10^6$\\
  \hline
  \# languages$_{> 0.0\%}$ & 19
  &
  \# languages$_{> 0.0\%}$ &  9 \\
  \hline
\end{tabular}
%% \begin{tabular}{|l|c|}
%%   \hline
%%   \multicolumn{2}{|c|}{Linux kernel \cite{LOClinux}} \\
%%   \hline
%%   \# LOC & $> 27 * 10^6$ \\
%%   \hline
%%   \# languages $> 0.0\% $ & 8  \\
%%   \hline
%%   \multicolumn{2}{|c|}{GNU C library \cite{LOCglibc}}\\
%%   \hline
%%   \# LOC & $\sim 1.9 * 10^6$ \\
%%   \hline
%%   \# languages $> 0.0\%$ & 9 \\
%%   \hline
%%   \multicolumn{2}{|c|}{{\tt Firefox} \cite{LOCfirefox}}\\
%%   \hline
%%   \# LOC & $> 44 * 10^6$\\
%%   \hline
%%   \# languages $> 0.0\%$ & 19 \\
%%   \hline
%%   \multicolumn{2}{|c|}{{\tt Blender3D} \cite{LOCblender}}\\
%%   \hline
%%   \# LOC & $5 * 10^6$\\
%%   \hline
%%   \# languages $> 0.0\%$ &  9 \\
%%   \hline
%% \end{tabular}
  \end{center}
\hrulefill
\end{table}

\paragraph{The size of real-world programs} %
%% The questions asked thus far about the material in 
%% \cref{fig:fact}, \cref{exa:fast-inverse-sqrt}, and
%% \cref{fig:flaky} were aimed at laying bare how difficult
%% it actually is to reason on the meaning of software.
%% Even three lines programs like the ones in \cref{fig:fact}
%% have non-trivial meanings.
Program size in reality differs dramatically from the one
of the material shown thus far. Consider the figures in
\cref{tab:software-size}.
Real-world software amounts to \millionS of lines of code
and is usually written in a variety of languages.

%% The linux kernel amounts to more than $27 * 10^6$ lines of
%% code (LOC),
%% \footnote{\url{https://seattlewebsitedesign.medium.com/what-systems-have-the-most-lines-of-code-and-why-48fd37d0af37}}
%% and of $22.6 * 10^3$ functions. 
%% %The kernel is but one part of an operating system,
%% thus the complexity does not end there.
%% The standard unix {\tt C} codebase, the {\tt glibc} library, amounts to $\sim 1.9 * 10^6$
%% LOC\footnote{\url{https://openhub.net/p/glibc/analyses/latest/languages\_summary}}
%% and contains $\sim 1.3 * 10^3$ functions\footnote{https://gist.github.com/PewZ/8b473c2a6888c5c528635550d07c6186}.
%Again this is but a part of an operating system, hence these figures
%are telling wrt the complexity of software artifacts.

The data in \cref{tab:software-size} is merely to a snapshot  at the time of writing
of the codebases of four well known software. One further and striking datum about
the Linux kernel is that a graphic driver alone exceeds $5.9$ \million LOC \cite{AMDdriver}.
To worsen the situation, software size appears to be ever increasing:
during his keynote at the $2005$ AOSD conference
\cite{boochKeynote2005}
\href{https://en.wikipedia.org/wiki/Grady_Booch}{Grady Booch}
answered the question
``how many lines of code are written each year around the world?'' thus
\begin{quote}
  about 35 \billion lines of source code are written each year. This is
based on about 15 \million or so software professionals world-wide, of
which current estimation is about 30\% actually cut code. [\ldots] 
%(The percentage of professionals cutting code has steadily declined :( ).
Each developer contributes about 7,000 LOC a year. Over history, we
have about a \trillion lines of code written so far!
\end{quote}

A suggestive depiction of how software size increases with time
is given in~\cite{linesofcode}, the sources of the data being at
the bottom of that web page.

The reader may argue that LOC is not a good metric about
software complexity. It turns out that ``no measure we presently
have is better at predicting effort or error rates than
simply counting the number of lines of code'' \cite{SEgreatestHits}.
Should the reader still prefer other metrics, let us consider the
number of functions in a given codebase. This metric seems to offer a
better granularity than LOC.
Consider the GNU C library \cite{glibc}, it contains $23.9 \times
10^3$ functions \cite{NOFglibc}.
Keeping in mind the precise semantics of {\em all} these 
functions, while it may be simpler than keeping in mind $\sim 1.9 \times 10^6$ LOC,
seems still no easy feat.

The sheer size of real-world programs that outlined above
calls for automatic program analysis. The next subsection
overviews briefly this matter.

%% MISINTERPRETATION OF THAT PAPER
%% It is not a coincidence that the semantics of the~$10$ lines of code
%% in \cref{exa:fast-inverse-sqrt} is hard to grasp. Consider for
%% instance that the brain of a young adult has a ``central memory
%% store'' limited to~$3$ to~$5$ symbols \cite{cowan2010}.
%Given this limitation,

\subsection{Automation}
Given the size of real programs, we would like to have tools that for any program $P$,
for every property $\phi$ of the behaviour of $P$
\begin{enumerate}[(1)]
\item\label{pt:exact-result} decide whether $P$ enjoys  $\phi$ or not,
\item\label{pt:auto} fully automatically,
\item\label{pt:finite} in finite time.
\end{enumerate}
{\em In general} these tools cannot exist.  %
Around 1936 in the papers that arguably started Computer Science as a discipline
\cite{Church1936-CHUAUP,DBLP:journals/x/Turing37}, \church and
\turing have proven that {\em in general} it is
impossible for a program (i.e. a tool) to decide whether
the execution of a program $Q$ terminates.
%The halting problem is {\em in general} undecidable.
In $1954$ Henry Gordon Rice has shown that our predicament is in fact worse:
{\em in general} every non trivial property of program
behaviour is undecidable~\cite{Rice1954-RICCOR-2}.

These impossibility results hold {\em in general}.
More often than not, though, program analysis has to be carried
out in very specific contexts, so the general limitations of
the techniques may not be a problem.
In other terms, given that \cref{pt:finite} above must hold,
different trade-offs between \cref{pt:exact-result} and \cref{pt:auto}
can be found, which suffice to meet practical purposes.
For instance, in spite of the undecidability results of
\cite{DBLP:journals/x/Turing37,Rice1954-RICCOR-2},
the halting problem is decidable for programming languages
expressive enough to write device drivers
\cite{DBLP:conf/cav/CookPR06}, and Turing machines with a lossy tape
\cite{DBLP:journals/iandc/AbdullaJ96}; moreover in languages
like \href{https://rocq-prover.org/}{Rocq},
recursion must be syntactically guarded,
which guarantees that programs always terminate.

Static type systems help finding nonsensical fragments of
programs (i.e. texts), thereby ensuring that
``well-typed programs don't go wrong'' \cite{DBLP:journals/jcss/Milner78}.
Arguably, type theory was conceived by
\href{https://en.wikipedia.org/wiki/Bertrand_Russell}{Bertrand
  Russell} exactly with this purpose \cite{Russell1908-RUSMLA},
because nonsensical fragments of propositions
are the ``reflexive fallacies'', i.e. the logical paradoxes, that type disciplines avoid:
\begin{quote}
  The division of objects into types is necessitated by the reflexive fallacies
  which otherwise arise. These fallacies, as we saw, are to be avoided by what may
  be called the " vicious-circle principle;" i. e., [\ldots]
  "Whatever contains an apparent variable must not be a possible value of that variable."
  [\ldots] This is the guiding principle in what follows.
\end{quote}

Not all programs that ``don't go wrong'' are well-typed. An example is
the program \mintinline{ocaml}{fun x -> x x}, so the type systems we
use in practice are sound but not complete. In theory this is a
limitation. In practice, though, this limitation does not hinder the
usefulness of type systems.

The need for practical tools and the tension between
\cref{pt:exact-result} and \cref{pt:auto} above have lead to the
development of different techniques for program analysis.
The authors of \cite{rival:hal-02402597} categorise them thus: 
\begin{inparaenum}[(a)]
\item testing,
\item assisted proving,
\item model-checking of finite state models,
\item model-checking at program level,
\item conservative static analysis,
\item bug finding.
\end{inparaenum}
We refer the reader to sections 1.3 and 1.4 of \cite{rival:hal-02402597} for a discussion
of the limitations and advantages of each one of these techniques and
we conclude this section hinting at a last practical trade off that
has been put forth by researchers in verification: {\em soundiness},
i.e. ``soundness, except for the usual caveats that experts know about
and consider inevitable.'' \cite{defenseunsoundness}.
In fact the practical need for
\begin{inparaenum}[(i)]
\item static analysis techniques that scale, and
\item engineering compromises
\end{inparaenum}
have also lead to the soundiness manifesto \cite{10.1145/2644805},
which advocates for analyses that are mostly sound (i.e. correct,
thanks to over-approximations of software behaviour), with specific
well-defined unsound choices (i.e. specific under-approximations of
software behaviour).
Since the use of {\em soundy} rather than {\em sound} techniques
for program analysis depends on the context (a graphical user
interface does not seem as critical as the software steering a
car), we think it practically useful to be aware of both approaches.

\begingroup
\begin{landscape}
\renewcommand{\arraystretch}{1.3}
%\newcolumntype{C}[1]{>{L\arraybackslash}p{#1}}
\begin{table}
  \centering
  \caption{Examples of incidents with a software root cause and significant, measured human or economic cost.}
  \label{tab:costs-potential-summary}
  \rowcolors{2}{lightgray}{white}
  \begin{tabularx}{\linewidth}{>{\raggedright\arraybackslash}p{2.7cm}lll>{\raggedright\arraybackslash}Xl}
    \toprule
    \rowcolor{white}
    Case & Sector & Human Cost & Economic Cost & Summary & Sources \\
    \midrule
    % Healthcare ---------------------------------------------------------------
    \hyperref[sec:costs-therac-25]{Therac-25} & Healthcare & Yes (6 deaths) & Unquantified & Race conditions in radiation 
    % therapy 
    software sent hazardous doses. & \cite{leveson1993therac, npr2011therac} 
    \\
    \hyperref[sec:costs-lascad]{London Ambulance} & Healthcare & Yes (20-30 deaths) & $1.5$ \millions \GBP & Ambulance dispatch system collapsed due to memory leak. & \cite{finkelstein1993report,fitzgerald2005lascad} 
    \\
    %\hyperref[sec:costs-carefusion]{CareFusion Alaris} & Healthcare & No & $244$ \millions \USD & TODO & \cite{bd2020alaris} 
    %\\
    % Engineering --------------------------------------------------------------
    \hyperref[sec:costs-sleipner-a]{Sleipner A Platform} & Engineering & No & $700$ \millions \USD & Modelling error in finite element analysis underestimated structural stresses resulting in the loss of the platform. & \cite{collins1997sleipner,jakobsen1994sleipner}
    \\
    % APIs --------------------------------------------------------------
%    \hyperref[sec:costs-cloudfare]{Cloudfare} & API & No & Unquantified & Flawed API implementation of resumption logic. & \cite{Cloudflare25,CloudflareVulnerability25} 
%    \\
%    \hyperref[sec:costs-drown]{DROWN} & API & No & Unquantified & Reuse of RSA private keys violating interaction protocol & \cite{Drown16,DrownSecurityAffairs16} 
%    \\
    % Aerospace ----------------------------------------------------------------
    \hyperref[sec:costs-patriot]{Patriot Failure} & Aerospace & Yes (28 deaths) & Unquantified & Clock drift due to truncation error prevented interception. & \cite{gao1992patriot} 
    \\
    \hyperref[sec:costs-patriot]{Boeing Starliner OFT-1} & Aerospace & No & $410$ \millions \USD & Synchronisation of spacecraft and launch vehicle clocks failed due to logical bug prevented planned orbital insertion. & \cite{lewis2020nasa,asap2020} 
    \\
    \hyperref[sec:costs-boeing-737]{Boeing 737 MAX} & Aerospace & Yes (346 deaths) & $20$ \billions\USD & Crashes due to single point of failure in fly-by-wire system. & \cite{faa2020tab,forbes2020boeing} 
    \\
    Mars Climate Orbiter & Aerospace & No & $327$ \millions \USD & Loss of mission due to unit mismatch (imperial vs.~metric) in flight control software. & \cite{nasa1999mco,nasa2009mco} 
    \\
    Ariane 5 Flight 501 & Aerospace & No & $370$ \millions \USD & Loss of mission due to float to integer conversion error in flight control software. & \cite{ariane501report,lelann1996ariane} 
    \\
    % Automotive ---------------------------------------------------------------
    \hyperref[sec:costs-toyota-sua]{Toyota SUA} & Automotive & Yes (89 deaths) & $1.2$ \billions \USD & Suspected software contribution to unintended acceleration. & \cite{BarrReport2013,toyotaKoopman2014} 
    \\
    %    \hyperref[sec:costs-tesla]{Tesla Autopilot} & Automotive & Yes (14 deaths) & Under investigation & Investigated for several fatal accidents involving software limitations. & \cite{} \\
    %    \hyperref[sec:costs-uber]{Uber Self-Driving} & Automotive & Yes (1 death) & Undisclosed & Failure to classify pedestrian due to software flaw. & \cite{} \\
    % Railways ---------------------------------------------------------------
    \hyperref[sec:costs-london-four-lines]{4LM Project} & Railways & No & $1$ \billion \GBP & Complexity of real-time software hindered development. & \cite{CostsLondonTube,TfLreport}
    \\
    % Accounting ---------------------------------------------------------------
    \hyperref[sec:costs-horizon]{Horizon IT Scandal} & Accounting & Indirect (13 suicides) & $1$ \billions \GBP (est.) & Software errors led to $900$+ false fraud accusations. & \cite{horizoncompensation,batesvspostoffice} 
      \\
    % IT Operations ------------------------------------------------------------
    \hyperref[sec:costs-crowdstrike]{CrowdStrike} & IT Operations & No & $8$ \billions \USD & Update crash led to global business disruptions. & \cite{crowdstrikePIR2019,nadella2019,crowdstrikeRootCause} 
    \\
    \hyperref[sec:costs-facebook]{Facebook DNS} & IT Operations & No & $150$ \millions \USD (est.) & Misconfiguration caused hours-long global outage. & \cite{CWFacebookDNS,MetaFacebookDNS} 
    \\
    \hyperref[sec:costs-eternalblue]{EternalBlue (WannaCry, NotPetya)} & IT Operations & Indirect & $14$ \billions \USD (est.) & Windows use-after-free vulnerability enabled remote code execution. NSA-developed exploit was leaked and used by ransomware and malware campaigns. & \cite{microsoft2017wannacry,wired2019eternalblue} 
    \\
    % Finance ------------------------------------------------------------------
    \hyperref[sec:costs-knight]{Knight Capital} & Finance & No &
    $460$ \millions \USD & Error in software operations. One server was not
    updated. &
    \cite{SEC315570,dseven2014,knightCapitalBBC}
    \\
    \hyperref[sec:costs-iron]{IRON} & Finance & No & $2$ \billions \USD (est.) & Infrastructure software contributed to market outage. & \cite{metltdowniron2023,wkIronFinance} 
    \\
    \midrule\rowcolor{white}
    \multicolumn{6}{p{\linewidth-\columnsep}}{
    Economic figures are reported as per the sources cited i.e., without any inflation adjustment or currency conversions, and labelled with (est.) when the source reports estimated costs.%, with \$ when in USD and with £ when in GBP.
    }\\
    \bottomrule
  \end{tabularx}
  \label{tab:examples-costs}
\end{table}
%% \begin{table}
%%   \centering
%%   \caption{Examples of software bugs which may lead to outages with known average costs.}
%%   \label{tab:costs-summary}
%%   \rowcolors{2}{lightgray}{white}
%%   \begin{tabularx}{\linewidth}{llll}
%%     \toprule
%%     \rowcolor{white}
%%     Case & Bug description & Potential Average Cost & Sources \\
%%     \midrule
%%     % APIs --------------------------------------------------------------
%%     \hyperref[sec:costs-cloudfare]{Cloudfare} &  Flawed API implementation of resumption logic.  & 11.1 \millions \USD &  \cite{Cloudflare25,CloudflareVulnerability25}
%%     \\
%%     \hyperref[sec:costs-drown]{DROWN} &  Reuse of RSA private keys violating interaction protocol & 8.01 + 8.85 \millions \USD &  \cite{Drown16,DrownSecurityAffairs16}
%%     \\
%%     \midrule\rowcolor{white}
%%     \multicolumn{4}{p{\linewidth-\columnsep}}{
%%     Economic figures are reported as per the sources cited i.e., without any inflation adjustment or currency conversions.
%%     }\\
%%     \bottomrule
%%   \end{tabularx}
%%   \label{tab:examples-potential-costs}
%% \end{table}
\end{landscape}
\endgroup

\section{High-stake failures and their costs}
\label{sec:costs}

In this section we present a series of real-world failures related
to software that are remarkable owing to their economical and/or human cost.
The prominent facts are summarised in \cref{tab:examples-costs}.
The list of software failures that we discuss in this
section is far from being exhaustive. In fact we have selected
%Among all the software failures we are aware of,
%we have picked the
the ones in \cref{tab:examples-costs} for we can
discuss them in our lectures, and they spur
interest in our students.
Our experience is that failures worth discussing 
with undergraduate and master students should have  
\begin{enumerate}
\item happened within the last $\sim 40$ years;
  the more recent the better, in particular if the students
  have actually experienced them;
  
\item an objective well-documented real-world cost
  (compensations for damages, casualties, \ldots);
% \end{enumerate}
% % MP: split the enumerate to avoid long white space before the table.
% \input{landscape_table}
% \begin{enumerate}
% \setcounter{enumi}{2} % continue from previous count
\item
  a root cause that has been identified and discussed in class:
  faulty code, poor software operations, poor software integration,
  exceeding software complexity, or lack of conformance with coding standards.
\end{enumerate}
Examples that do not respect the above metrics are here because of
historical interest, or because of the sheer cost of the failure.
\cref{tab:examples-costs} reports the Mars Climate Orbiter and the
Ariane $5$ accident for historical reasons, but we do not discuss them.
%In particular, academics have been citing the Ariane 5 for decades with the
%result (perhaps among other ones) of making it bore students to death.
In particular, the Ariane 5 accident has been thoroughly studied
and clear accounts are given in the official report \cite{ariane5flight501failure}
and the talk \cite{talkAriane5}. The reader interested in this matter may also
watch the footage of the actual accident \cite{ariane5footage}.

%%
%% TWO FAILRUES TO DISCUSS FOR THE SECOND ROUND
%% 
%% Two additional software failures that we do not discuss are the
%% Northeast blackout of $2003$ \cite{wkNortheasternBlackout} and
%% the Lufthansa Flight 2904 accident in $1993$ \cite{wkLufthansa}.

\subsection{Healthcare}

\parcase{Therac-${\mathbf{25}}$}{sec:costs-therac-25}
The Therac-$25$ was a computer-controlled radiation therapy machine
developed by Atomic Energy of Canada Limited (AECL) in the early
$1980$s. 
Between $1985$ and $1987$, six known patients in the United States and
Canada received massive overdoses of radiation -- some more than $100$
times the intended dose -- resulting in severe injuries and at least
six confirmed deaths \cite{leveson1993therac, npr2011therac}.

The root cause of these incidents was a race condition in the
code that allowed the machine operators to enter commands in
rapid succession, causing the system to bypass safety checks.
%%% ORIGINAL
%% The root cause of these incidents was a failure in the machine's software. 
%% Unlike its predecessors, the Therac-$25$ relied heavily on software to
%% manage safety interlocks and beam control. 
%% A race condition in the code allowed operators to enter commands in
%% rapid succession, causing the system to bypass safety checks.
Specifically, if treatment settings were modified quickly via the
terminal, the software could misconfigure the beam and deliver a
high-powered electron dose without the protective target in place
\cite{leveson1993therac}.
The software lacked proper synchronisation for concurrent processes,
and key hardware safety features had been removed or disabled under
the assumption that the software alone would prevent dangerous
configurations.
The Therac-$25$ relied heavily on software to manage safety interlocks
and beam control, unlike its predecessors.

These technical %failures
issues were compounded by poor development practices: there was no
independent review of the code, testing was inadequate, and AECL
did not act promptly on early warnings.
Crucially, no \formalverification or systematic analysis of the
safety-critical code was performed. 

Although detailed financial figures have not been made public, the
incident led to legal settlements, reputational damage for AECL and
the hospitals involved, and long-term mistrust in the safety of
computerised medical devices.
The human cost was tragically clear: at least six patients died and
many more suffered irreversible harm due to preventable software errors.

Further details can be found in \cite{DailyWTFtherac}, which presents
an interesting account and a detailed analysis of the Therac-$25$
failure.

\parcase{London Ambulance Service}{sec:costs-lascad}
In \date{10}{1992}, the London Ambulance Service (LAS) deployed a new
Computer-Aided Dispatch (CAD) system designed to automate the
allocation of emergency medical resources across Greater London.
Within hours, the system began to fail catastrophically, leading to
severe delays in ambulance response times including life-critical
cases \cite{finkelstein1996lascad, wilson1994lascad,
  fitzgerald2005lascad, darren99}.

The system was developed under intense time pressure and tight
budgets, and it exhibited severe architectural weaknesses.
It could not cope with high volumes of real-time events, failed to
degrade gracefully under load, and lacked robust error handling.
Moreover, it ran on outdated hardware that was unable to support the
demands of the software
\cite{finkelstein1993report,dillon1993lascad}.
Organisational issues compounded the technical faults: the LAS
underestimated the difficulty of automating previously manual
procedures, and failed to anticipate the operational risks of rapid
deployment in a safety-critical environment.

The immediate technical cause of the collapse was a memory leak in a
small portion of code \cite{dillon1993lascad}.
Incident data remained in memory even after it was no longer needed,
eventually exhausting available resources and causing system-wide
failure.
Such a defect could have been uncovered through systematic stress
testing or the application of \formalverification techniques suited to
critical systems.

The financial impact included the write-off of approximately \money{\GBP}{1.5}{\million}
in public funds allocated to the development and
deployment of the system, widespread operational disruption and the
associated additional operational costs, and reputational damage to
the LAS and its contractors.
The human impact of the delays in administering emergency care is
difficult to assess.
Multiple sources report fatalities attributable to these delays in the
20-30 range \cite{fitzgerald2005lascad,independent1992lascad,independent1992lascad}.

%
%\parcase{CareFusion Alaris Infusion Pump}{sec:costs-carefusion}
%

\subsection{Engineering}

\parcase{Sleipner A Platform}{sec:costs-sleipner-a}
% \gb{Marco: We should state explicitly that the problem was due to how the software was used, and not to the software itself.
%   This may let us stress that (for instance) not only software has to be correct, but also that validation of its results must be carefully carried out;
%   or that anyhow the principle Garbage In Garbe Out always holds.
%   This seems in line with the conclusion of the analysis that you already wrote.
% }
%
On \date[23]{12}{1991}, the concrete gravity base structure of the Sleipner A offshore platform collapsed during a controlled submersion test in the Gandsfjord near Stavanger, Norway. 
The platform, owned by Statoil (now Equinor), sank rapidly and was completely
destroyed. 
Although no lives were lost, the financial loss was estimated at around \money{\USD}{700}{\million} \cite{lindberg2022sleipner,collins1997sleipner, melchers2001sleipner, jakobsen1994sleipner, arnold2008sleipner}.

\begin{figure}[t]
  \begin{tabular}{ccc}
    \centering
    %[trim={left bottom right top},clip]
    \includegraphics[trim={7mm 7mm 1mm 6mm},clip,width=.24\textwidth]{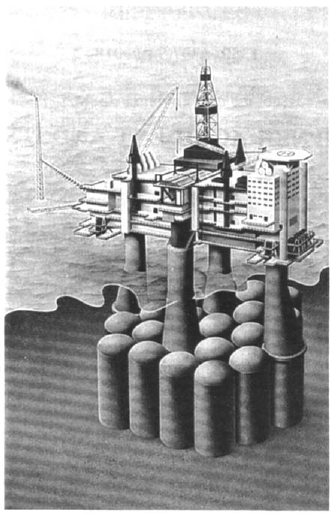} &
    \includegraphics[width=.34\textwidth]{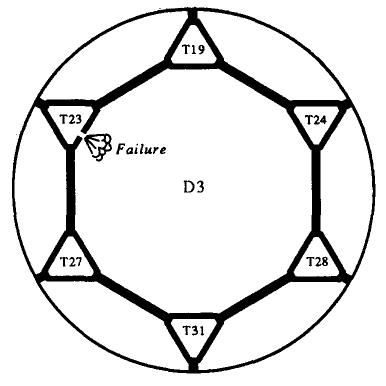} &
    \includegraphics[width=.37\textwidth]{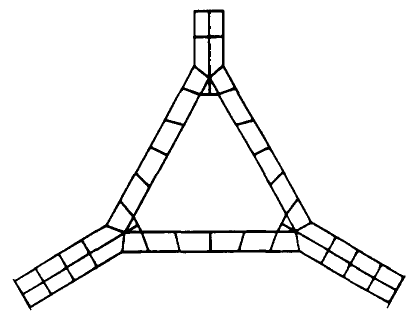} \\
    (a) Sleipner A Platform & 
    (b) Tricell wall where failure occurred. &
    (c) Tricell finite element model.
  \end{tabular}
  \caption{The Sleipner A Platform structural failure (cred.~\cite[Fig.~1,3,4]{jakobsen1994sleipner}).}
  \label{fig:sleipner}
  \hrulefill
\end{figure}

An extensive investigation identified compounding causes rooted in how structural simulation software was used during the platform's design \cite{arnold2008sleipner,soereide1997sleipner,dnv1991sleipner,holand1997sleipner}. 
The engineering team employed NASTRAN, a general-purpose finite
element analysis (FEA) tool, to model stress distribution in the
concrete structure.
FEA works by discretising a structure into smaller components -- a
mesh of finite elements -- and approximating how each responds to physical forces.
The investigation determined the discretisation used for a critical
component the platform's ballast chambers, the \emph{tricell}, was too coarse
and contained some skewed\footnote{The \emph{skewedness} is a critical mesh
quality metric that measures how much a finite element deviates from
its ideal shape. Highly skewed cells significantly deviate from the
ideal shape potentially leading to numerical instability and inaccurate
results.} elements (cf.~\cref{fig:sleipner}) and thus the analysis
underestimated the shear forces in the concrete walls of the tricell
by ~$45\%$.
The results of the engineering team have been checked by an
independent team using a different FEA software, SESAM.
However, the discretisation used by this verification team was too
coarse to characterise local defects like the one in the tricell walls
\cite{soereide1997sleipner}.
As a result, the final design assumed lower forces in critical parts
of the ballast than the ones it would actually experience, and thus
when the platform was submerged during testing, the under-reinforced
wall cracked, water flooded in, and the structure disintegrated.

The Sleipner A collapse is now widely cited as a cautionary example of the risks associated with relying on numerical simulations without sufficient verification and attention to the measures of their quality. 
While the finite element software produced stable and plausible outputs, these were based on incorrect modelling assumptions that went unchecked \cite{jakobsen1994sleipner,arnold2008sleipner,soereide1997sleipner}. 
This highlights the broader challenge of ensuring that complex models accurately reflect the real-world systems they aim to represent and of navigating the trade-offs between the cost and precision of models.
In this context, \formalmethods--such as specification checking,
constraint validation, or the use of formally verified modelling
components--could have served as an additional safeguard.
By helping to specify and verify the conditions under which particular
modelling techniques are valid, \formalmethods can reduce the risk of
critical misapplications and support more robust validation practices
in safety-critical engineering.

\subsection{Aerospace and Defence}

% TODO: short paragraph about aerospace examples. 
% Ariane 5 Flight 501 \cite{ariane501report,lelann1996ariane}
% Mars Climate Orbiter \cite{nasa1999mco,nasa2009mco}

% Lufthansa Flight 2904

\parcase{Patriot Missile Failure}{sec:costs-patriot}
On \date[25]{2}{1991}, during the Gulf War, a United States Army
barracks in Dhahran, Saudi Arabia, was struck by an Iraqi Scud
missile.
The Patriot missile defence system, which was designed to intercept
incoming missiles, failed to engage the target.
As a result, the Scud struck the barracks directly, killing $28$
American soldiers and injuring around $100$ others
\cite{gao1992patriot}.

\begin{figure}[t]
  \sidecaption % SPRINGER, up to 7.8cm 
  \includegraphics[trim={20mm 3mm 0mm 2mm},clip,width=7.8cm]{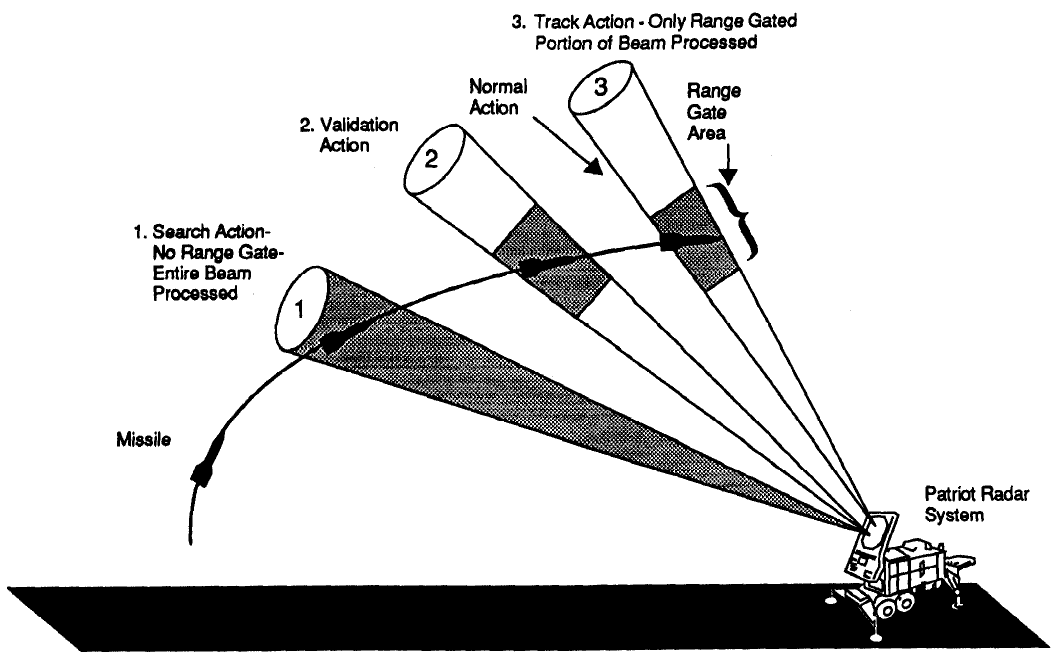}
  \caption{The range gate is a specific segment of time during which the system anticipates receiving a reflection of a radar signal from a target based on the time it takes for a radar signal to travel to the target and return to the radar receiver. 
  By calculating this round-trip time, the system can determine the distance to the target and predict where the target should be during the next radar scan. 
  % The range gate is crucial for tracking and engaging fast-moving
  % targets, as it allows the system to focus on specific areas in space
  % where the target is expected to be, thereby improving accuracy and
  % reducing computational load.
  % Therefore, accurate timekeeping is crucial for the system to track high-speed targets.
  (Cred.~\cite[Fig.~3]{gao1992patriot}.)}
  \label{fig:patriot}
  \hrulefill
\end{figure}

The root cause of the failure was traced to a software defect
in the Patriot missile system's internal time-keeping
mechanism. Specifically, the system used a $24$-bit fixed-point
register to store the time in tenths of a second since system startup,
and performed time calculations by multiplying this value by $0.1$.
The longer the system remained operational, the more floating-point
rounding errors accumulated and the more inaccurate its internal
clock became.
At $8$ hours of operation, the error was $20$\% of the range gate (cf.~\cref{fig:patriot})
% MP: footnote explaining the range gate in place of the figure. 
% \footnote{The range gate is a specific segment of time during which the system anticipates receiving a reflection of a radar signal from a target based on the time it takes for a radar signal to travel to the target and return to the radar receiver. 
%   By calculating this round-trip time, the system can determine the distance to the target and predict where the target should be during the next radar scan. 
%   % The range gate is crucial for tracking and engaging fast-moving
%   % targets, as it allows the system to focus on specific areas in space
%   % where the target is expected to be, thereby improving accuracy and
%   % reducing computational load.
%   % Therefore, accurate timekeeping is crucial for the system to track high-speed targets.
% }
of the Patriot system \cite{crichton2012patriot,ding2025machine}.
By the time of the attack, the system had been running continuously
for over $100$ hours, causing a time drift of approximately $0.34$
seconds -- an interval sufficient for the fast-moving Scud missile to be
mislocated by several hundred metres \cite{gao1992patriot}.

% REPEATED
% 
% This failure illustrates how seemingly minor software assumptions can
% lead to catastrophic outcomes when left unchecked in safety-critical
% systems. The Patriot system was not originally designed for continuous
% operation over such extended periods, and the software had not been
% rigorously validated to operate under these conditions.

This failure illustrates how seemingly minor software assumptions can
lead to catastrophic outcomes when left unchecked in safety-critical
systems.
The Patriot missile system was not originally engineered for prolonged
continuous operation, and its software had not been rigorously
validated under such extended uptime conditions.
The resulting timing drift -- though technically small -- became
significant enough over time affect the effectiveness of this critical
automated defence system.

%A fix for the issue had been identified shortly before the
%incident but constraints of the battlefield operations hindered its
%deployment.
%\todo{So what ? This is orthogonal to the description of the failure.}

% MP: I wasn't able to find information on whether a workaround like "please reboot the system every X hours" was issued or if the producers downplayed the issue so I avoided speculating. Nonetheless, I thought this higlighs the importance of maintenance/operations. I have expanded this point into the following but if this is too long or enlarges the scope it can be left out without problems. 

% The consequences were grave: the loss of $28$ lives and a blow to confidence in automated defence systems.

The incident prompted serious scrutiny of software validation
practices within military systems and is now widely cited as an
example of the dangers of ignoring numerical precision errors and
time-dependent faults in real-time systems \cite{postol1991patriot,ding2025machine}.
% TOO Detailed
%
% \footnote{
% For instance, the incident is cited as one of the early cases in the proposal in the context of military systems of the Software Development Lifecycle Theory which ``expands the causal timeline of accidents beyond decisions made on the battlefield to those made decades earlier in software design''
% \cite{ding2025machine}.}
%
Notably, a software fix for the timing issue had already been identified shortly before the incident.
Two weeks before the incident, Army officials received Israeli data
indicating $20$\% loss in accuracy after the system had been running
for $8$ consecutive hours and quickly modified the software to improve
the system's accuracy \cite{gao1992patriot,crichton2012patriot}.
However, the realities of combat operations -- including logistical
constraints, mission priorities, and the inherent difficulty of
updating software in the field -- delayed its deployment.
This highlights a broader lesson: even well-validated systems can
remain vulnerable if maintenance and patching procedures are not
realistically integrated into their operational context.
In safety- and mission-critical environments, ensuring that updates
are feasible, timely, and secure is just as important as writing
correct software in the first place.

\parcase{Boeing Starliner OFT-1}{sec:costs-starliner}
On \date[20]{12}{2019}, Boeing's CST-100 Starliner spacecraft launched on
its first uncrewed Orbital Flight Test (OFT-1) as part of NASA's
Commercial Crew Program. Although the launch vehicle, an Atlas V
rocket, successfully placed Starliner into space along the planned
trajectory, a series of software anomalies prevented the spacecraft
from reaching its intended orbit and docking with the International
Space Station (ISS).
The mission was terminated early, and the capsule returned safely to
Earth landing at the White Sands Missile Range two days
later \cite{lewis2020nasa,nasaspaceflight2020}.

The root cause of the failure was a defect in the software that
manages Starliner's Mission Elapsed Timer (MET), specifically how it
is synchronised with the other key systems of the mission.
Starliner's flight software was designed to synchronise with the Atlas
V's timing system during \emph{terminal count} -- the final phase of
the countdown just before launch -- when all systems are expected to
synchronise.
However, the software instead retrieved the timestamp from the Atlas V
much earlier, at a point when the launch vehicle's internal timer had
not yet been reset for the mission \cite{lewis2020nasa,asap2020}.
This resulted in the MET reporting a time 11 hours ahead of the actual
mission time which induced the flight control software to initiate
inappropriate thruster firings that consumed excessive fuel,
ultimately preventing the rendezvous with the ISS.
Later investigation revealed that the software \enquote{lacked that terminal
count requirement} \cite{spacenews2020}.

A second critical software issue, identified during the mission,
involved incorrect mapping of separation commands between the crew
module and its service module.
Had this fault not been discovered mid-flight -- where it was
mitigated by a last-minute software patch uploaded from the ground --
it could have caused a collision during module separation.
NASA and the Aerospace Safety Advisory Panel (ASAP) concluded that
this defect could have resulted in the loss of the spacecraft, and
possibly of crew, had the mission been crewed
\cite{asap2020,lewis2020nasa}.

These issues were attributed to insufficient end-to-end software
integration testing and a weak safety culture in Boeing's development
process.
In its public recommendations, the ASAP expressed \enquote{concern
  with the rigor of Boeing's verification processes}, and called for
stronger validation and verification practices \cite{asap2020}.
NASA's Office of Inspector General estimated the failed mission cost
nearly \money{\USD}{410}{\million}, as a second uncrewed test flight
(OFT-2) had to be scheduled before any crewed flights could proceed
\cite{spacepolicy2019}.

\parcase{Boeing 737 MAX}{sec:costs-boeing-737}
The Boeing 737 MAX was grounded worldwide in \date{3}{2019}
following two catastrophic crashes -- Lion Air Flight 610
in \date{10}{2018} and Ethiopian Airlines Flight 302 in \date{3}{2019}
-- that together claimed the lives of 346 people.
The incidents triggered a global crisis of confidence in Boeing and
the (US) Federal Aviation Administration (FAA), leading to intense
scrutiny of the aircraft's design, certification, and regulatory
oversight processes. The financial impact was also severe.
Already in 2020 Boeing estimated it had incurred costs related to this
crisis for approximately \money{\USD}{18.4}{\billion}, including
compensation to airlines, production halts, legal settlements, and the
cost of software fixes and re-certification efforts
\cite{reuters2020boeing,forbes2020boeing}.
The reputational damage to Boeing and FAA regulators remains
long-lasting, and the crashes led to major reforms in US aircraft
certification policies.

At the centre of both crashes was the Maneuvering Characteristics Augmentation System (MCAS), a software-based control system introduced to mitigate pitch-up tendencies caused by design changes in the 737 MAX's engine placement.
MCAS was designed to automatically lower the aircraft's nose when it detected a high Angle of Attack\footnote{
The Angle of Attack is the angle between the chord line of an aircraft's wing -- a straight line connecting the leading and trailing edges -- and the vector of the relative wind, or the direction of the airflow encountered by the aircraft. 
% TOO DETAILED
%
% AoA is crucial as it directly influences the aerodynamic lift generated by the wings. 
% At low angles, airflow remains smooth over the wing surface, optimizing lift production. 
% As the AoA increases, lift also increases up to a critical threshold. 
% Beyond this critical angle, airflow becomes turbulent and separates from the wing surface, drastically reducing lift and resulting in an aerodynamic stall. This principle is vital for flight control systems -- like the Boeing 737 MAX's MCAS -- which rely on accurate AoA measurements to maintain aircraft stability and safety.
} (AoA). 
Critically, it relied on input from a single AoA sensor at a time. 
When that sensor failed—providing falsely high readings—MCAS repeatedly activated, forcing the aircraft into a nose-down attitude despite pilot attempts to regain control \cite{faa2020tab,ntsb2021boeing}.

Although MCAS performed as designed, the root issue lay in its flawed design and integration. 
The MCAS software had unchecked authority over flight control surfaces overriding pilot inputs without them being notified of the MCAS engagement.
It lacked redundancy in sensor input, meaning that a single sensor failure could render the system effectively `blind'.
It was neither adequately disclosed in pilot training nor documented in flight manuals.
% TOO DETAILED
%
% \footnote{
% Following the Lion Air crash, Boeing issued a bulletin describing procedures to counteract MCAS activation. 
% Ethiopian Airlines pilots had received this guidance. 
% During Flight 302, the faulty AoA sensor fed erroneous data to MCAS, which repeatedly activated. 
% Although the pilots attempted to follow the recommended procedures -- including disabling the electric trim system -- MCAS had already commanded nose-down trim multiple times. 
% %The resulting configuration of control surfaces, combined with high speed and altitude loss, made recovery physically unachievable. 
% The final investigation concluded that by the time manual trim was attempted, aerodynamic forces on the stabiliser were too great to be countered by manual effort alone \cite{ntsb2021boeing}. 
% }
Furthermore, Boeing's internal safety analyses underestimated MCAS's risk profile and did not fully consider failure scenarios.
These oversights were compounded by regulatory lapses, as the FAA had delegated significant portions of the aircraft's certification process to Boeing itself \cite{faa2020tab,house2023boeing,doj2021boeing}.

The 737 MAX disasters exemplify a systemic failure in software-based system design, verification, and certification in a high-assurance domain. 
In this case, the issue was not a coding error per se, but a breakdown in safety culture, requirements engineering, and independent verification of software behaviour under fault conditions that did not prevent critical but avoidable single point of failures.
Under stronger safety culture, \formalmethods could have contributed to the specification and verification of safety-critical behaviours, particularly with respect to redundancy requirements, failure modes, and pilot interfaces.

\subsection{Transportation: automotive and railways}

\parcase{Toyota SUA}{sec:costs-toyota-sua}
The Toyota Sudden Unintended Acceleration (SUA) incidents occurred over the period between $2008$ and $2011$\cite{ToyotaRecall11,ToyotaSUA13}.  
Toyota vehicles were reported to be accelerating without driver input, leading to accidents and fatalities. 
Although initial investigations by the National Highway Traffic Safety Administration (NHTSA) and the National Aeronautics and Space Administration (NASA) suggested that mechanical issues such as "sticking" accelerator pedals and floor mat entrapment was causing the high-speed unintended acceleration, subsequent analysis concluded that the car malfunction was in fact caused by a software error~\cite{DeptOfTransToyota19}.

Software experts uncovered significant issues within Toyota's Electronic Throttle Control System (ETCS) software~\cite{BarrReport2013,toyotaKoopman2014}. 
A 2013 court investigation concluded that the source code of the ETCS was of "unreasonable quality," containing bugs that could lead to the unintended acceleration. 
Experts pointed out that the software architecture allowed single-point failures, lacked adequate fail-safes, and had a complex, convoluted structure described as "spaghetti code"~\cite{ToyotaSpaghetti13}. 
These deficiencies meant that certain faults could disable safety mechanisms without alerting the driver (\emph{e.g.,} single-bit corruptions in RAM could disable safety mechanisms without setting off diagnostic trouble codes, and a stack overflow in one of the ETCS tasks could cause the software task responsible for throttle control silently failed without triggering any warning), increasing the risk of uncontrolled acceleration.

The root causes of these software failures were attributed to Toyota's internal development practices. 
Numerous investigations reported that the software department of the company did not adhere to established best practices for safety-critical software development. 
In particular, they identified issues such as inadequate code review processes, insufficient testing and other forms of verification, and a lack of proper software documentation.
For instance, Toyota failed to perform any \formalverification or adequate testing on the ETCS software. There was no evidence of sufficient code coverage metrics being used leaving many software paths untested, nor did the company stress-test the software for fault conditions, which is a standard requirement for safety-critical embedded software.
Additionally, the corporate culture of the company was accused of prioritising rapid development and cost-saving over rigorous safety verification, leading to systemic oversights in software quality assurance.
Toyota's development practices did not align with known standards such as MISRA-C (a coding standard for safety-critical automotive software in C) or ISO 26262 (the international standard for functional safety of automobile electronic systems). 
Experts concluded that poor software design (\emph{e.g.,} lack of abstraction with over 80,000 global variables) significantly increased the potential for data corruption and race conditions.
Moreover defensive programming techniques essential for embedded systems, such as input validation, redundancy and fail-safe mechanisms were either missing or poorly implemented.

The repercussions of these software errors extended beyond the tragic loss of lives and numerous injuries.
The company faced substantial legal and financial penalties where it had to settle hundreds of lawsuits related to SUA incidents, amounting to significant financial costs~\cite{ToyotaLawsuit13}. 
The scandal tarnished the reputation of the company and prompted a reevaluation of software safety standards across the automotive industry. 
It underscored the critical importance of rigorous software verification and validation processes, especially as vehicles become increasingly reliant on complex electronic systems.

\parcase{The Four Lines Modernisation Project}{sec:costs-london-four-lines}
The transport project for London’s Four Lines Modernisation (4LM) program aimed to upgrade the four Sub-Surface train lines with new signalling and rolling stock. 
Initially, a \money{\GBP}{354}{\million} signalling contract was awarded to Bombardier in 2011, but it was terminated in 2013/2014 due to inadequate progress and performance concerns. 
In 2015, Thales was awarded a revised signalling contract worth approximately \money{\GBP}{760}{\million}.
At the time \money{\GBP}{85}{\million} was paid to Bombardier to exit the failing contract \cite{CostsLondonTube}, of which \money{\GBP}{67}{\million} was considered a direct waste, as there was no deliverable received for that portion of the payout \cite{TfLreport}.

Central to the delays and contractual failure were software issues mainly due to incomplete or incorrect logic, poor integration between subsystems, and insufficiently mature control-software modules~\cite{Reichl:TAP:2016}. 
Documents and commentary highlighted that Bombardier struggled with delivering a fully working, integrated solution.  
Defects in critical movement-authority logic, fail-safe state-transition handling at junctions, and overlapping service-control interfaces, manifested during late-stage testing and simulations, which meant that the system failed to meet defined safety and performance requirements. 
The complexity of timing-critical software interactions and inconsistent assumptions between functional modules further exacerbated verification challenges, undermining confidence in the software quality and repeatability.
Railway standards such as EN 50128 / IEC 62279 strongly recommend the use of formal specification and verification methods for signalling works, which can expose critical logic flaws in the early stages of development~\cite{ItehStandard:2015}. 
The recommended formal modelling of signalling control and fallback recovery procedures (\emph{e.g.} via state-machines) would have revealed any inconsistent transitions or missing guards that later contributed to integration failures.
Moreover, formal data validation and property-based testing of interlocking route tables, timetable interactions, and timing constraints would have helped catch mis-configurations or timing anomalies, enhancing test coverage in complex software subsystems.

%While there were no reported injuries,
The impact of the software-related delays were significant: the cancellation of the Bombardier contract and the re-letting to Thales increased cost substantially (from \money{\GBP}{354}{\million} to \money{\GBP}{760}{\million}). 
Transport for London estimated that the overall programme authority grew to approximately \money{\GBP}{5.4}{\billion}, with \money{\GBP}{1}{\billion} of additional cost arising from overruns and failures relating to the first contract, as well as several years of delay. 
These postponements deferred anticipated benefits (such as capacity improvements, reliability gains, better train frequency) and necessitated extended periods of service interruption.

\subsection{Information Technology}
\parcase{Horizon IT scandal}{sec:costs-horizon}
The Horizon system was introduced in $1999$ by Fujitsu (successor of
ICL) to digitise UK Post Office accounting and till operations,
becoming one of Europe's largest non-military IT
projects~\cite{wallis2025great}.
The system, used for accounting and stock-taking,
inaccurately recorded losses and money missing in branches. 
Immediately after rollout, subpostmasters reported discrepancies such
as unexplained shortfalls, frozen screens, and duplicate
transactions—bugs that were consistently dismissed by management.

Between $1999$ and $2015$, over $900$ subpostmasters faced criminal
prosecutions based solely on Horizon-generated shortfall data.  There
were  $700$ convictions, leading to $236$
imprisonments~\cite{horizoncompensation,justiceforpostmasteralliance,horizoninquiry2025}.
The fallout included bankruptcies, loss of livelihoods, family
breakdowns, severe mental distress, 
%and at least four confirmed suicides.
and four confirmed suicides, which may actually be thirteen \cite{suicides}.
By $2024$, the scandal had become the largest miscarriage of justice in modern British history, prompting a publicly funded compensation scheme for over $4,000$ affected individuals, totalling more than \money{\GBP}{1}{\billion}.
In a video was posted on the official youtube channel of 10 Downing Street, the then Prime Minister Rishi Sunak opened his speech thus: \enquote{For decades hundreds of subpostmasters were wrongfully convicted as a result of faulty post-office IT. Lives were destroyed, reputations ruined, and innocent people left fighting for justice.} In the same speech Prime Minister also remarked that \enquote{this has been one of the greatest miscarriages of justice
in our nation's history.} \cite{sunak2024}.
As of \date[31]{1}{2025}, approximately \money{\GBP}{663}{\million} had been paid to over $4.300$ claimants \cite[Section
  2.2]{horizoncompensation}.
At the time of writing $10.000$ people seek compensation \cite{suicides}.
Victim compensation has become a legislative matter \cite{horizoncompensationact2024}: the economical compensation raised questions for it relies on taxpayers money~\cite{taxpayers2025}, and at least one MP suggested Fujitsu should pay compensation to victims of its faulty IT system~\cite{fujitsucaoughup2024}.

Technical investigations revealed hundreds of unresolved bugs in Horizon, such as the `Dalmellington' freezing bug and the `Callendar Square' duplication bug. 
The software suffered from poor coding quality and inadequate testing, with Fujitsu and Post Office staff having knowledge of dozens of serious defects since as early as $1999$~-~$2001$. 
Compounding the issue, the system's design lacked sufficient controls to prevent or detect user errors and system anomalies, violating basic principles of defensive design in safety-critical systems.

The root cause of the failures stemmed from a combination of factors.
First, ICL/Fujitsu’s management favoured rapid deployment over rigorous quality assurance, accepting Horizon despite internal criticism and lack of developer capability.
\cite[Appendix B, pag. 287]{originasofadisaster2022} describes the
attitude of the company thus:
\enquote{ICL's whole attitude to delivery of the project has been consistently
  to find reasons not to do what is asked, but to cut corners: this
  does not bode well for a reassuring view of the future \ldots}.
Second, Post Office executives repeatedly dismissed warnings from subpostmasters, legal appeals, and internal audit reports—choosing to prosecute individuals rather than investigate the system~\cite{postofficecorporate2025}.
Third, government oversight failed to challenge assumptions of
accuracy, even after MPs raised numerous concerns in $2009$, which
indicated systemic negligence.  

Collectively, responsibility lay not only with Fujitsu's defective
software engineering but also with the Post Office's refusal to
acknowledge faults, and regulatory bodies' failure to intervene.

Given the jaw dropping scale of the Horizon scandal, we gather
further material in \cref{sec:horizon-appendix}.

\parcase{CrowdStrike Falcon update}{sec:costs-crowdstrike}
On \date[19]{7}{2024} 
% at $04:09$ UTC, 
CrowdStrike issued a routine `rapid response' configuration
update for the Windows sensor to gather telemetry on possible novel
threat techniques \cite{crowdstrikePIR2019}.
Within minutes, an update misalignment triggered \millionS of Windows machines worldwide to display the infamous Blue Screen of Death (BSOD).
At $05:59$ PM, the CEO of Microsoft posted the following tweet:
\enquote{Yesterday, CrowdStrike released an update that began impacting IT
systems globally. We are aware of this issue and are working closely with
CrowdStrike and across the industry to provide customers technical
guidance and support to safely bring their systems back online.}\cite{nadella2019}.
%
% \href{https://en.wikipedia.org/wiki/Troy_Hunt}{T. Hunt} responded thus
% ``This is basically what we were all worried about with Y2K, except
% it's actually happened this time \includegraphics[height=8px]{images/skull.png}``
%
CISA (Cybersecurity and Infrastructure Security Agency), along with national cybersecurity agencies in the UK, Australia, and Canada, confirmed the incident was a faulty update—not a cyberattack~\cite{CloudStrikeFalconCISA24}.
Correspondents from prominent news organisations called the events that followed this update the \enquote{largest IT outage in history} \cite{cnbc2019,hunt2019}.

The software update brought down~$\sim 8.5$ \million machines world wide,
and affected more than~$19$ countries \cite{cybersecuritynews2019}.
According to \cite{ukgovt2024} the  Crowd Strike incident
affected $60$\% of GP practices with disruption to software holding
patients’ appointment details, prescriptions, and health records
showed the catastrophic impact of IT and cyber threats on people's lives.
The outage total cost to the UK economy was estimated
between~\money{\GBP}{1.7}{\billion} and \money{\GBP}{2.3}{\billion}
(\money{\USD}{2.18}{\billion} and~\money{\USD}{2.96}{\billion})
\cite{kovrr2024}, and the estimated loss in revenue for the 
top~$500$ US companies, excluding Microsoft, is nearly~\money{\USD}{5.4}{\billion}
\cite{wkCrowdStrike}.
At the time of writing the legal and economical repercussions on CrowdStrike
are still on-going and both the SCC and the US Justice Department are corresponding with CrowdStrike \cite{madmoney2025}.

According to the External Technical Root Cause Analysis \cite[pag. 11 and 12]{crowdstrikeRootCause}  the root cause of the bug was
a an iteration over $21$ entries, performed on an input array
of only $20$ pointers. 
Since memory area supposed to be the
$21^{\mathit{st}}$ entry of the array contains an invalid pointer,
the program attempts a read through the invalid pointer,
resulting in an out-of-bounds read.

Note that according to the \href{https://cwe.mitre.org}{MITRE} Common Weaknesses
Enumeration (CWE)  Out-of-bounds Read is the $6^{\mathit{th}}$ most
common CWE in $2024$ \cite{cwe2025}.
The incident resulted thus from key software malpractices: lack of strict
input validation, absence of buffer checking, poor content validation,
and inadequate testing procedures for updates that interact deeply
with the OS kernel.

% This striking software failure can be summarised thus:
% in $2024$ an \underline{out of bounds read} costed to the UK and USA
% economies more than some $7$ \billions USD.
% Any CS student should let this sentence sink in.

About a year later, during the writing of this chapter, an incident with a similar failure mode resulted in another (albeit more limited) global outage. 
On \date[12]{7}{2025}, a policy update to Google Cloud's Service Control triggered widespread 503 errors for users of Google's core cloud services and Workspace products \cite{google2025cloud}. 
Google's incident report identifies the root cause in a memory handling fault without proper error handling introduced by a recent binary release.
The report observes that \enquote{the code update and binary release went through [their] region by region rollout} without issue because
\enquote{the code path that failed was never exercised during this rollout due to needing a policy change that would trigger the code}.
% The root cause of this outage mirrors the failure mode in the CrowdStrike incident: a fault without proper error handling in a code path that was enabled by the policy global roll-out and that had not been properly tested snowballed into a systemic failure.
To Google's credit, the update included a `red-button' (a procedure for quickly rolling it back) and this enabled recovery to begin in less than an hour and the incident to be globally resolved within a few hours. 
While no numeric cost estimates are available at the time of this writing, the similarity in root cause with CrowdStrike and the global scale of Google infrastructure and services highlight the high stakes of identical software vulnerabilities in large-scale software updates.

\parcase{Facebook DNS Table Update Failure}{sec:costs-facebook}
On \date[4]{10}{2021} Facebook and its affiliated services,
including Instagram, WhatsApp, Messenger, and Oculus, experienced a
global outage that rendered the platforms inaccessible to \billionS of
users worldwide lasting approximately six to seven
hours~\cite{WikipediaFacebookDNS,CWFacebookDNS}. 
The outage was traced back to a loss of IP routes to Facebook's Domain
Name System (DNS) servers, which were self-hosted. 
The withdrawal of these routes from the Border Gateway Protocol (BGP) 
prohibited users from resolving Facebook and related domain
names, thus removing Facebook's presence from the Internet's navigation system~\cite{GuardianFacebookDNS}.

The root cause of the outage was attributed to a configuration error
during routine maintenance~\cite{MetaFacebookDNS}.
A command intended to assess the global backbone network capacity
inadvertently disconnected all of Facebook's data centres from the
Internet.
Facebook's DNS servers operated on a separate network for safety
reasons, they were configured to withdraw their BGP routes whenever
they could not connect to the data centres.
This setup led to the DNS servers becoming unreachable, which further
compounded the outage.

Although the potentially harmful command was part of a routine
maintenance procedure within the company, it lacked adequate
safeguards to prevent such a widespread
disconnection~\cite{CloudfareFacebookDNS,MetaFacebookDNS,GuardianFacebookDNS}.
It exposed a single point-of-failure design flaw in the system,
together with tight coupling between unrelated components (routing and
DNS) which lead to cascading failure.
Moreover, once the network collapsed, Facebook engineers were locked
out of the internal recovery systems (that were hosted on the same
infrastructure that went offline), suggesting a lack of redundant,
out-of-band access channels (e.g., separate secure tunnels or backup
control systems).
The situation was exacerbated by the company's reliance on its own
services for internal communication and access, hindering their
ability to respond swiftly to the crisis.
More specifically, there was no external fallback tooling, such as
third-party monitoring, DNS services, or cloud management backups;
this violates the requirements of ISO/IEC 12207 (Software Lifecycle
Processes) for controlled and traceable software
changes~\cite{RegisterFacebookDNS}.
In addition, the absence of robust validation mechanisms allowed the
erroneous command to propagate, highlighting deficiencies in
Facebook's software verification processes: a well-engineered system
would typically use pre-deployment verification, staging environments,
and automated rollback mechanisms.
This violates IEEE 1012 (Software Verification and Validation
Standard) due to the lack of \formalverification of configuration
scripts, the missing regression testing or staging simulation and
independent validation of high-risk
changes~\cite{RegisterFacebookDNS,CloudfareFacebookDNS}.

The outage impacted the operations of the company resulting in financial losses. 
Internally, Facebook employees were unable to access essential tools and facilities, including email, security badges, and communication platforms.
Externally, the outage disrupted services for \billionS of users and businesses worldwide, with an estimated loss of at least \money{\USD}{60}{\million} in advertising revenue~\cite{WikipediaFacebookDNS}. 
The incident underscored the critical importance of rigorous software verification, robust fail-safes, and comprehensive disaster recovery planning in large-scale network operations. 

On \date[14]{7}{2025}, Cloudflare suffered a global outage of its 1.1.1.1 DNS service that lasted approximately 62 minutes. 
A configuration update to a non-production `Data Localization Suite' inadvertently removed the BGP announcements\footnote{BGP announcements are control messages used in the Border Gateway Protocol (BGP), the routing protocol that governs how packets are exchanged between autonomous systems on the Internet. 
These announcements advertise the availability of IP address prefixes and inform neighbouring networks about how to reach them. 
%Withdrawing an announcement makes a prefix unreachable globally, effectively rendering the associated service inaccessible from the Internet.
} for Cloudflare's resolver prefixes, effectively pulling the service offline worldwide \cite{cloudflare2025dns}.
Although the full impact analysis is still pending, the root cause, a dormant misconfiguration in service topology, mirrors the underlying pattern seen in the Facebook incident. 
This recurrence reflect deeper, systemic risks in managing critical network infrastructure at scale.

\parcase{EternalBlue}{sec:costs-eternalblue}
On \date[14]{4}{2017}, a group known as the Shadow Brokers leaked a set of cyberweapons developed by the U.S.~National Security Agency (NSA), including a tool known as EternalBlue \cite{liu2022eternalblue,ars2017eternalblue,wapo2017eternalblue,wired2019eternalblue,guardian2017eternalblue}. 
EternalBlue targeted a use-after-free memory bug in Microsoft's implementation of the SMBv1 network protocol to enable a remote code execution on Windows systems affected by this vulnerability (CVE-2017-0144 \cite{CVE20170144}). 
The NSA held on this vulnerability for more than five years and alerted Microsoft only in \date{2}{2017}---Symantec later found that the vulnerability and possibly parts of the NSA toolkit had been obtained by Chinese groups by \date{2}{2016} \cite{wired2019eternalblue,guardian2017eternalblue}.
Microsoft released a critical patch (MS17-010) in March, a month before EternalBlue was publicly leaked.
However, the widespread delay in applying the patch allowed the
exploit devastating effective, and it was quickly adopted by many
malware campaigns
\cite{liu2022eternalblue,tp2017retefe,nyt2017notpetya,wired2019eternalblue}.

In \date{5}{2017}, the WannaCry ransomware worm used EternalBlue to infect over 300,000 computers across 150 countries in a matter of hours; 
accepted conservative estimates place the global economic impact of WannaCry around the \money{\USD}{4}{\billion}  mark \cite{cbs2017wannacrycost,loop2024stats}.  
Security researcher Marcus Hutchins discovered a `kill switch' in the malware: registering a specific domain for a DNS sinkhole stopped the attack because the worm would encrypt the victims file only if it was unable to reach a hardcoded domain \cite{liu2022eternalblue}.
In the following days, the kill-switch domain was flooded by a barrage of distributed denial-of-service (DDoS) attacks, yet Hutchins' intervention provided defenders precious time to deploy mitigations and patch systems. 
Compared to attacks of the same type, the impact of WannaCry could have been much worse had its developers not planted kill switch or had they specifically targeted highly critical infrastructure \cite{guardian2017eternalblue}.

On \date[27]{6}{2017}, the anniversary of the Ukrainian constitution, the NotPetya malware used EternalBlue causing economic damages in excess of \money{\USD}{10}{\billion} \cite{wired2018notpetya,loop2024stats}. 
The attack started in Ukraine with an update of Ukrainian tax accounting system and although it eventually spread to other countries, $80\%$ of the affected systems were Ukrainian, primarily ministries, state-owned businesses, and critical infrastructures like railways, water plants, and power plants including nuclear ones like the monitoring system of the Chernobyl nuclear power plant.
The aims of the attack were to cause damage and disrupt the Ukrainian State rather than economic gain as with WannaCry: NotPetya posed as a variant of the ransomware Petya but was in fact a wiper, a type of malware designed to wipe data and cause unrecoverable damage to the infected systems. 
Forensic and geopolitical analyses strongly suggest perpetrators affiliated to the Russian State \cite{nyt2017notpetya}.

The EternalBlue incident demonstrates how a single unpatched vulnerability in ubiquitous software can quickly escalate into a systemic global crisis. 
It further underscores the importance of \formalverification of protocol implementations, timely and verifiable patch deployment, and system-level threat modelling in critical infrastructure. 
In particular, the vulnerability was rooted in semantics of memory management and pointer use, domains where \formalmethods can play a key role in prevention and certification.
% a vulnerability formally modelled later in academic work using Meta Attack Language \cite{novokhrestov2024creating}

This incident confronts our society with the consequences of a government decision to hide information about a `weapon-grade' vulnerability that could cripple critical infrastructure, disrupt businesses, and harm its citizens.
In the wake of the WannaCry outbreak, Microsoft President Brad Smith publicly criticised the hoarding of cyberweapons by nation-states arguing that that governments must take more responsibility for the vulnerabilities they discover and stockpile \cite{microsoft2017wannacry}.
In his widely cited statement, he provides the following striking comparison to frame the danger posed by leaks like this,
\begin{quote}
\enquote{An equivalent scenario with conventional weapons would be the U.S. military having some of its Tomahawk missiles stolen.}
\end{quote}
Following the incident, many called stronger international accords and bodies to regulate cyberweapons and protect civilians from digital fallout, emphasising the shared responsibility of both governments and businesses in global cybersecurity.
Although these ethical questions are beyond the scope of this chapter, we believe they are too important for our field and society to be omitted.

\subsection{Finance}

\parcase{Knight Capital Group}{sec:costs-knight}
On \date[1]{8}{2012}, Knight Capital Group, one of the
largest U.S. equity market makers, deployed a routine update for its
automated trading system, SMARS, and experienced a significant error
in the operation of its automated routing system~\cite{SEC315570}.
The update incorporated new logic for RLP orders that inadvertently
reused a deprecated `Power Peg' flag, which had not been fully removed
from its code.
This setting activated a dormant test module not intended for production, setting the stage for chaos.
%
% Knight Capital was an American global financial services firm. The
% following material essentially en excerpt of Administrative
% Proceeding of the Securities And Exchange Commission.
%

Within 45 minutes of the market opening, SMARS executed approximately
$4$~\million unintended trades, handling nearly $397$~\million shares
across $154$ stocks~\cite{KnightCaptial19}.
Owing to this error, incurred a pre-tax loss of around \money{\USD}{440--460}{\million}
over a~$45$ minute period, wiping out roughly $75\%$ of its market value in two days.
The firm required an emergency capital injection and was eventually
acquired, ending its run as an independent entity.

A combination of systems faults and procedural failures triggered the debacle.
Apart from the legacy code still present in production software
(`Power Peg' was deprecated but never removed),  there was a faulty
deployment process where one of eight servers did not receive the
updated code (this procedure was not automated but done manually), and
the deployment script failed silently.
The software also exhibited lack of input validation and safety
checks, since it had no safeguards to detect or restrict anomalous
trading loops.
Finally, inadequate (regression) testing, no kill switch, and no
independent verification (governance) meant that the catastrophe could
not have been prevented or, at least, contained.

Knight Capital’s software failure was not just a coding error, but a
systemic breakdown of software engineering principles~\cite{cisq2022}.
The incident underscores why compliance with standards like IEEE 1012
(Verification and Validation) and ISO/IEC 12207 (Software Life Cycle)
are critical for high-stakes financial systems.
%
% If these frameworks had been followed, the catastrophe likely would have been prevented or at least contained.

% To enable its customers’ participation in the Retail Liquidity Program (“RLP”) at the New York Stock Exchange, Knight made a number of changes to its
% systems and software code related to its order handling processes.
% These changes included developing and deploying new software code in
% SMARS. SMARS is an automated, high speed, algorithmic router that
% sends orders into the market for execution.

% Beginning on \date[$27\th$]{July}{2012}, Knight deployed the new
% RLP code in SMARS in stages by placing it on a limited number of
% servers in SMARS on successive days. During the deployment of the new
% code, however, one of Knight's technicians did not copy the new code
% to one of the eight SMARS computer servers. Knight did not have a
% second technician review this deployment [\ldots].
% Knight had no written procedures that required such  a review.

% The root cause of the failure is a human error, and a poor update
% process. According to \cite{cisq2022} this is an example of
% {\em operational failure}. We added this example to this section as a
% reminder that not only software has to be correct, but that it also
% has to be operated via correct business processes.

\parcase{IRON}{sec:costs-iron}
Iron Finance was a decentralised finance (DeFi) protocol that launched
the IRON  algorithmic {\em stablecoin} on the Polygon blockchain.
The value of the IRON coin can be seen on the coinbase web-site \cite{ironvalue}.
%\footnote{https://ironfinance.medium.com/}

A stablecoin is cryptocurrency designed to maintain a stable value
relative to specific assets.
According to \cite{metltdowniron2023} the idea was for the stability
of IRON to work as follows:
partially collateralise the stablecoin with USDC — it was around $75$\%
for most of its short lifespan — with the rest coming form a volatile
asset, TITAN, which backed the remaining $25$\% of the remaining value.
In turn USDC and TITAN are respectively
(a) a cryptocurrency pegged to the United States
dollar \cite{wkUSDC} and (b) the cryptocurrency used in
the IRON Finance DeFi.

The stability was to be ensured thus: If the price of IRON fell
below $1$ USD, arbitrageurs could buy it for less then $1$ USD on the market,
and then redeem it with the IRON treasury (a contract on the Polygon
chain) for $0.75$ USDC and $0.25$ worth of TITAN \cite{metltdowniron2023}. 

In \date{6}{2021} a rapid increase in TITAN's value triggered a
large-scale sell-off by major holders, and this caused TITAN's value
to collapse to nearly zero and IRON to lose its dollar peg \cite{wkIronFinance}.
Note now that the following line of code of the IRON smart contract,
\begin{minted}{C}
require(_share_price > 0, "Invalid share price");
\end{minted}
where \mintinline{C}{_share_price} refers to the price of TITAN, as
provided by an oracle that is reporting it as $0$.
Since this line of code is in the \mintinline{C}{redeem} function of
the IRON smart contract and the condition
\mintinline{C}{_share_price > 0} evaluates to false due to rounding,
the USDC on the blockchain {\em cannot} be redeemed, and thus
\money{USDC}{272}{\million} are still locked in the contract \cite{metltdowniron2023}.

The root cause of the IRON fiasco is the difference between a strict
comparison, i.e.~\mintinline{C}{>}, and a non-strict
one,~\mintinline{C}{>=}.

The Iron Finance debacle sent shock waves through the decentralised
finance (DeFi) sector, and it serves as a cautionary tale about the
inherent risks and complexities of algorithmic stablecoins and the
DeFi ecosystem \cite{di2024blockchain}.
Further information on the TITAN crash can be found in the paper \cite{breakapeg2024}.

\section{Potential costs of software bugs}
\label{sec:potential-costs}

Software validation in general, and formal verification in particular,
are aimed at producing software that works correctly, and that may
not cause any failure.
It is thus important to invest in validation and verification
{\em before} any damage is done. How much to invest?
A natural manner to answer this question is via the {\em potential costs}
that bugs may cause. %
To make this point we summarise in
in \cref{tab:examples-potential-costs}
two bugs whose potential costs amount to millions USD, and
we discuss both bugs and potential costs in the rest of this section.
%\cref{sec:API-bugs}.

\begingroup
%\begin{landscape}
\renewcommand{\arraystretch}{1.3}
%\newcolumntype{C}[1]{>{L\arraybackslash}p{#1}}
\begin{table}
  \centering
  \caption{Examples of software bugs which may lead to outages or data breaches with known average costs.}
    \label{tab:examples-potential-costs}
  \rowcolors{2}{lightgray}{white}
  \begin{tabularx}{\linewidth}{llll}
    \toprule
    \rowcolor{white}
    Case & Bug description & Potential Avg. Cost & Sources \\
    \midrule
    % APIs --------------------------------------------------------------
    \hyperref[sec:costs-cloudfare]{Cloudfare} &  Flawed API implementation of resumption logic.  & 11.1 \millions \USD &  \cite{Cloudflare25,CloudflareVulnerability25}
    \\
    \hyperref[sec:costs-drown]{DROWN} &  Reuse of RSA private keys violating interaction protocol & 16.86 \millions \USD &  \cite{Drown16,DrownSecurityAffairs16}
    \\
    \midrule\rowcolor{white}
    \multicolumn{4}{p{\linewidth-\columnsep}}{
    Economic figures are reported as per the sources cited i.e., without any inflation adjustment or currency conversions.
    }\\
    \bottomrule
  \end{tabularx}
\end{table}
%\end{landscape}
\endgroup

%\subsection{Communication Platforms and API Infrastructure}
%\label{sec:API-bugs}
\parcase{Cloudflare's Mutual TLS vulnerability}{sec:costs-cloudfare}
Cloudflare Mutual TLS (mTLS)~\cite{Cloudflare25} is a two-way authentication layer for software parties in communication over the Transport Layer Security (TLS) protocol.
The client (user/device) and server verify each other's identities using digital certificates before any data is exchanged.
This adds a security layer for APIs, IoT, and Zero Trust environments to prevent spoofing, on-path attacks, and unauthorised access by verifying certificates issued by trusted Certificate Authorities (CAs). 
It allows non-login requests (like from IoT devices) to prove their legitimacy and strengthens internal security by verifying all internal service-to-service communication. 

In late January 2025, Cloudflare was alerted to a vulnerability in how its mTLS implementation handled TLS session resumption~\cite{CloudflareVulnerability25}. 
Due to a flaw in session resumption logic, a client with a valid mTLS certificate for one Cloudflare “zone” could reuse a resumed TLS session ticket to access another zone without needing to revalidate its certificate for that second zone.
More concretely, when a session was resumed for a different Cloudflare zone, the BoringSSL TSL library implementation reused the cached certificate and assumed its verification status was still valid, skipping full certificate validation for the new zone.
This bypassed the expected certificate authentication, potentially allowing unauthorised access where mTLS was assumed to be enforced.

Several established software engineering practices might have prevented or mitigated this vulnerability. 
Rigorous protocol modelling and secure design reviews of the session resumption paths could have identified that cached client authentication state might be applied incorrectly across security contexts. 
Isolation testing and fuzzing of session cache boundaries — especially for security-critical features like mTLS — would likely have surfaced improper reuse of authentication state. 
The project should have used automated security testing in CI/CD pipelines and implemented stronger assertions and defensive runtime checks to ensure that resumed sessions for mTLS always re-validate certificates against the full trust chain when required, which would have caught the core vulnerability pre-release.

Cloudflare tracked the issue as CVE-2025-23419 and mitigated it by disabling session resumption for mTLS within about 32 hours of notification, with no evidence found of active exploitation. 
At the time of writing Cloudflare reported no evidence of active exploitation and there is no official account of the costs incurred.
However industry research suggests that security incidents and
certificate-related outages can be expensive when they cause service
disruption or downtime~\cite{Axelspire25} %
%% (\emph{e.g.} certificate
%% outages cost \money{\USD}{11.1}{\million} each on average whereas
%% compliance failures cost \money{\USD}{14.4}{\million} each on
%% average).
(\emph{e.g.} on average each certificate
outage costs \money{\USD}{11.1}{\million},
and on average each compliance failure costs \money{\USD}{14.4}{\million}).
Since mTLS is used in high-security API environments,  bypassing certificate validation would expose sensitive back-end services to unauthorised access. 
There were certainly operational costs associated with emergency mitigation (disabling session resumption and rolling out the fix) and engineering time spent analysing and hardening the codebase. 
From a business perspective, such vulnerabilities would affect reputation and customer confidence, especially for clients relying on zero-trust security models. 
%
% Moreover, handling disclosures and coordinating fixes consumes resources and can lead to regulatory or compliance scrutiny depending on industry sectors customers operate in.

\parcase{DROWN Cross-Protocol TLS/SSLv2 vulnerability}{sec:costs-drown}
The DROWN (Decrypting RSA with Obsolete and Weakened eNcryption) vulnerability was publicly disclosed in March 2016 after a team of international researchers found that attackers could decrypt modern TLS sessions by exploiting weaknesses in the obsolete SSLv2 protocol (CVE-2016-0800)~\cite{Drown16}. 
It can affect servers that support SSLv2 or share RSA private keys with other services that do, even if the primary service uses modern TLS. 
The researchers showed that by capturing enough encrypted traffic and then making crafted connections to an SSLv2-enabled server, an attacker could recover session keys and read supposedly secure communications that used up-to-date TLS protocols. 
At the time of disclosure, approximately 33\% of HTTPS servers were vulnerable because of lingering SSLv2 support or key reuse across services~\cite{DrownSecurityAffairs16}.

The root cause of DROWN was twofold: the continued support for the long-deprecated SSLv2 protocol and the reuse of RSA private keys across SSLv2 and modern TLS services. 
SSLv2 was known to be insecure since the 1990s and formally deprecated long before the DROWN disclosure, yet many server configurations and cryptographic libraries still permitted SSLv2 connections for backward compatibility. 
Because SSLv2 contained cryptographic weaknesses — particularly in how it handled export-grade ciphers and RSA padding — attackers could leverage it as a cross-protocol oracle to decrypt traffic encrypted under TLS. 
Even servers that did not directly support SSLv2 could be vulnerable if they used the same private key as another SSLv2-enabled service, such as an email server.

Engineering practices such as protocol deprecation management that track and phase out support for obsolete protocols such as SSLv2 should be built into secure software lifecycle processes, with clear sunset timelines and automated checks to prevent unsupported versions from being enabled. 
Key management hygiene, such as unique RSA keys per service, would have prevented an SSLv2 service from undermining the security of TLS services. 
More importantly, comprehensive security verification and auditing, including both static analysis of a formalised protocol model as well as dynamic scanning for legacy protocol support, could have flagged SSLv2 support and key reuse during development or deployment. 

To date, there is no public record of a specific financial figure tied to fixing DROWN across all affected systems. 
Nevertheless, industry research at that time (\emph{i.e.} in 2016) 
costed data breaches at around \money{\USD}{400}{} per record\footnote{A record describes one individual data subject’s information entry that contains sensitive or protected data (\emph{e.g.} a customer profile, a patient entry or a user account).}~\cite{ponemon2016cost}.
In addition, the average cost of identifying date breaches was estimated at \money{\USD}{5.83}{\million} when the mean time to identify a data breach was less than 100 days, and \money{\USD}{8.01}{\million} otherwise.  
The broader economic impact of the vulnerability is substantial and includes the investment required for patching and disabling SSLv2 across millions of servers, the operational costs of updating cryptographic libraries and configurations, potential compliance-related expenses, and the reputational and risk mitigation costs that organisations face when legacy protocol support exposes user data.
Data breach containment was costed at \money{\USD}{5.83}{\million} when it was achieved within 30 days, and \money{\USD}{8.85}{\million} when it took longer~\cite{ponemon2016cost}. 
%
% The widespread nature of the issue — with potentially millions of HTTPS servers at risk — underscores how technical debt in cryptographic protocols can translate into real-world cost and risk.

%\section{Successful cost mitigation}
\section{Return of investment from \formalverification}
\label{sec:advantages}
\label{sec:mitigation}
In general \formalmethods, i.e. rigorous mathematical techniques,
are so useful to society that in our history some mathematical techniques
have been rediscovered more than once.
Consider these two examples.
Leibniz and Newton in the 17th century introduced calculus, and in particular
the definite integral \cite{wkIntegral},
i.e. a technique to compute the area underneath a curve.
In 1994 the paper \cite{10.2337/diacare.17.2.152} proposed a
similar idea, using a technique that actually dates back to 350BC \cite{wkTaiModel}.
A second example is the back propagation algorithm, which is at the heart
of neural networks \cite[Chapter 6]{10.5555/574634}. It is essentially
an efficient application of the chain rule proposed first by G. W. Leibniz in
1676 \cite{wkBackPropagation}.

In the rest of this section, we review the available evidence that
using rigorous mathematically-founded techniques for \formalverification
of software can lead to an overall cost mitigation and to a more
sustainable development process (both economically and otherwise,
also environmentally).

There is a rich body of case studies documenting the evidence of the
application of \formalmethods in industrial practice
\cite{WoodcockEtAl2009,BicarreguiEtAl2009,KulikEtAl2022,LeComteEtAl2017,GnesiEtAl2012,terBeekEtAl2024};
we also refer to thirty years of the Proceedings of the International
Conference, formerly Workshop, on \FormalMethods for Industrial
Critical Systems. Additionally, there have been secondary studies
distilling quantitative evidence of successful application
\cite{Fitzgerald2013,Linger1993,OsaiweranEtAl2013}. Without trying to
being exhaustive we decided to review three examples of different
types of \formalmethods applied to two different domains, for which
there is empirical evidence of cost mitigation and more sustainable
development.

\subsection{Healthcare} 
Healthcare has been one of the primary application domains of
\formalmethods; we refer to some surveys
\cite{BockenholtEtAl1992,Bonfanti2018} for an overview of the
application of \formalmethods in the healthcare domain. Instead of
compiling a full list of success stories, we focus on one case, where
ample empirical evidence shows the positive impact of the use of
\formalmethods in significantly improving the quality of the product
and reducing the product costs.  

Analytical Software Design (ASD) \cite{Broadfoot2003,Hopcroft2004} is
a tabular representation of state machines, developed by a Dutch
company called Verum. The corresponding commercial tool ASD:Suite
translates such tabular specifications into Communicating Sequential
Processes \cite{BrookesEtA1984}. This allows for validation and
verification against basic properties such as deadlock-freedom, as
well as composition with monitors and verifying more domain-specific
properties.   

Philips Healthcare started applying ASD in its software development
process, particularly for the Interventional X-Ray Machines. 
There,
ASD was applied to various sub-components of the front-end- and
back-end subsystems \cite{OsaiweranEtAl2010}. 
Statistical data were gathered and analysed \cite{GrooteEtAL2011,OsaiweranEtAL2016} comparing the quality of
components with various degrees of ASD-generated quote and the results
suggest a significant improvement in quality and effort wherever ASD
is used for both sub-systems. In particular, for the back-end code,
they found that the ASD-generated code has a four-fold better quality
(in terms of defects per line of code) compare to hand-written
code. Also they report a three-fold increase in productivity (measured
in the effective lines of code) by using the ASD method.

\subsection{Railways} 
There is a rich history of application of \formalmethods in the
railway industry; we refer to  surveys
\cite{BasileEtAl2018,Fantechi2014,FerrariEtAl2022,terBeekFG25} and
overview papers \cite{Amendola2020,LeComte2025}. Most of the
references provided in this section are taken from a recent survey by
ter Beek et al. \cite{terBeekEtAl2024}, where more examples of the
application of \formalmethods in railway (and otherwise) are provided.  

For example, B-method has been extensively used in various railway
projects, including in Automatic Train Protection (ATP) system
\cite{BehmEtAl999,DaSilva1992,Guiho1990,Essame2006} employed in
several busy railways lines with outstanding results. Also a popular
urban-signalling system has been developed using the B-method
\cite{ComptierEtAl2019}, which is now in operation in more than 100
metro lines world-wide.
Due to these success stories, the use of \formalmethods is enshrined
in railways standards; for example, for critical functions (at Safety
Integration Levels 3 and 4) the European standard CENELEC EN 50128
\cite{CENELC-EN-5018} for railway software recommends the application
of \formalmethods.

In the context of the European 4SECURail
project \cite{4SECURailwebpage} an economical cost-benefit analysis of the adoption of \formalmethods
has been performed.
The Benefit/Cost Ratio (BCR), indicative the amount of savings divided by the amount of costs are calculated as a metric for the financial gain of applying formal methods; 
for this cases, the BCR is calculated at 5.05 \cite{BelliEtAl2023}, 
meaning that the adoption of
\formalmethods is projected to provide a five-fold economical benefit
compared to its costs. 

\subsection{Aerospace and defence}

Like for healthcare and railways, aerospace and defence have also been
a historical area of application for \formalmethods due to the high
risks and certification requirements. 
We refer the interested readers to the surveys
\cite{cofer2012,AitAmeur2009,Zhao2017} and discuss a few recent
high-profile cases.

% Airbus
Airbus replaced traditional unit testing of safety-critical avionics
code with \formalverification (`unit proof' in their lingo) under the
avionics certification DO-178B/C, using theorem proving and model
checking to establish correctness of low-level requirements starting
with the programs for the A400M, A380, and A350 \cite{yannick2013}. 
The adoption of CompCert at Airbus we reported in the \hyperref[sec:introduction]{Introduction} is an example of this move \cite{airbusWCET}.
Overall, the unit proof initiative cut verification time and costs by
reducing manual test development and execution and while Airbus has
not publicly released any financial estimates, industry surveys
indicate typical cost savings of $30-50\%$
\cite{cofer2012,yannick2013}.

% Frama-C @ Dassault
Dassault-Aviation is another firm that gainfully employs \formalverification during the development and certification of their avionics software. 
For instance, during a pilot program in 2004, a team of engineers
working on the flight-control software for the Flacon jet used Frama-C
to replace integration robustness testing and verified.
Already 20 years ago, Frama-C was able to handle $85\%$ of the
requirements assertions and analyse code blocks of 50 KLOC on average. 
They report that \enquote{because robustness verification is a
  recurrent task, the gain is roughly a person-month per flight
  software release} \cite{yannick2013}.

% Collins Aerospace
Collins Aerospace (Rockwell Collins) adopted \formalverification to address the cost of DO-178B/C certification. 
Specifically, they used open source, SMT-based model checking tools to
automate the peer review and test generation activities during the
development of critical avionics software for crew alerting systems,
flight controls, and synoptics systems \cite{wagner2019}.
These are large avionics systems with thousands of test cases and reduction in manual test case writing can provide savings in the range of hundreds of person-days \cite{wagner2019}.
Similarly, Rolls-Royce employed the C Bounded Model Checker, an
automated, open-source model checker for ANSI-C, to generate test
cases for safety-critical avionics software in alignment with
DO-178C/ED-12C standards~\cite{youcheng2017}.

\subsection{Information technology}

Formal verification is traditionally associated with safety- or
mission-critical domains such as aerospace, defence, railways, and
nuclear systems. However, it is increasingly being applied beyond
these areas, driven by improvements in tooling, education, and
lightweight techniques, as well as by the escalating costs of IT
failures illustrated in \cref{sec:costs}.
Here we report some illustrative examples of this trend. 

% TLA+ @ espark
At eSpark Learning, engineers applied TLA+ for just two days to model
a critical service migration and, besides identifying serious bugs
that prevented the loss of a major client, the team cites the
formalisation as key to avoiding approximately USD 100k per year in
additional maintenance costs~\cite{espark2020}.

% TLA+ @AWS
Amazon has been increasingly applying \formalverification across its products since at least 2011 \cite{brooker2025,Newcombe2015}.
For instance, between 2011 and 2014 TLA+ has been applied in over 10
complex, high-scale, critical systems at AWS including S3, DynamoDB
and EBS volumes and consistently uncovered subtle bugs and enabled
aggressive performance optimisations with confidence
\cite{Newcombe2015}.
In one case, modelling a replication protocol traced a 35-step concurrency bug that had slipped through QA and code reviews; fixing it before implementation saved approximately two months of development time on a four-month project which translates to hundreds of thousands of dollars in avoided rework.
The report remarks a low cost of adoption saying that
\enquote{Engineers from entry level to principal have been able to
  learn TLA+ from scratch and get useful results in two to three
  weeks}, which is in contrast to the common assumption that effective
use of \formalverification is too costly outside of safety-critical
industries.
% HOL Light @ Amazon
In a more recent instance, a team at Amazon's Automated Reasoning used
the HOL Light proof assistant to formally verify and optimise RSA
signature routines on the Graviton2 CPU achiving achieved a throughput
boost of between $33\%$ and $94\%$, depending on key size
\cite{lee2024}.
% TODO: P @AWS
% TODO: Dafny @AWS

% Allow @ Rackspace
The cloud provider Rackspace applied the Alloy model checker to an
existing production system and uncovered a design flaw severe enough
to require a year's worth of team effort to correct
\cite{rackspace2019}.

% Frama-C @ Thales
Java Card is a technology that enables Java applications on smart cards and is found in critical application including SIM, payment, identity, and authentication cards. 
In 2021, Thales used Frama-C to mathematically prove security properties of over 7,000 lines of C code in its JavaCard virtual machine checking over 52.000 conditions. 
This effort enabled Thales to achieve the world's first EAL7 Common Criteria certification for such a system while eliminating extensive manual evaluation and testing. 
% TLA+ smart card

% CI/CD
%Formal verification tools are increasingly more common in CI/CD pipelines. 
For instance, \cite{Howard2025} binds a formal TLA+ specification to a
real C++ implementation of Microsoft's Confidential Consortium
Framework and integrate verification into the CI pipeline uncovering
six subtle bugs before release
paper. This hybrid approach of melding lightweight formal methods with
automated testing illustrates how formal verification is becoming
practical and scalable even in modern, evolving IT systems. 

%% Intel https://dl.acm.org/doi/pdf/10.1145/1391469.1391675

\section{Establishing a Mindset}
\label{sec:beyond}

\Formalmethods, with their emphasis on unambiguous language and
rigorous reasoning grounded in mathematics, offer lasting value beyond
software correctness guarantees.
We argue that this mindset enhances education in areas beyond
verification and that it mitigates technical debt by helping
preserve institutional memory\footnote{\emph{Institutional memory}
refers to the collective knowledge and learned experiences held by an
organisation, which are preserved through artefacts, processes, and
documentation, enabling continuity despite personnel changes.}.

\paragraph{Education}
There are established curricula and materials, and  ongoing research
and discussion on teaching \formalmethods
e.g.,~\cite{Ferreira2024FMTea,FMTea2019,FMTea2021,FMTea2023,TFM2004,TFM2009,Ahrendt2009}.
In particular, we invite the reader to consult the
website \cite{fmetea} of the Formal Methods Teaching Committee of \FormalMethods Europe.
Beyond being a subject in its own right, \formalmethods used in
software verification can play a valuable role in supporting the
teaching and learning of other subjects.

Formalising a theory not only sharpens its logical structure but also
provides insights into how the theory may be presented more
effectively. Some educators report that the process of formalisation
helps identify conceptual bottlenecks that students struggle to
grasp. In \cite{CMP23}, for example, the authors recount their
experience formalising a theory of Choreographic Programming while
teaching it in a concurrency theory course for postgraduate computer
science students at the University of Southern Denmark. They note
that:
\begin{quote}
\enquote{It quickly became apparent that the technical aspects [\dots]
  that complicated the formalisation of [Choreographic Programming]
  were also the most challenging for the students.}
\end{quote}
Furthermore, the goals of making a theory more intuitive for students
and amenable to machine encoding can reinforce one another. The same
authors observe that:
\begin{quote}
\enquote{The seemingly disparate goals of making the theory more
  intuitive to students and amenable to formalisation actually
  converged on the same solution, and sometimes resulted in useful
  exchanges of feedback.}
\end{quote}
In the case above, students benefited indirectly from the
formalisation even when they were not exposed to it directly.

Many educators now integrate proof assistants into courses where
\formalverification is not the focus and where students are not
typically introduced to such tools.
For instance, in \cite{JV2022,FVB2020}, the authors describe their use
of Isabelle/HOL in an introductory logic course at the Technical
University of Denmark. A student survey revealed that they felt
\enquote{formalisations helped them understand concepts in logic, and
  that they experimented with them as a learning tool.}
More broadly, a recent survey \cite{TGN2025} highlights their growing
presence of proof assistants across undergraduate logic, mathematics,
and computer science courses, aided by enhanced interfaces and
tailored input languages.
Indeed, usability and learnability—often cited as barriers—have been
receiving more attention and resources, reducing the cost of
adoption.
For instance, \cite{HLY24,M24} advocate for the use of Lean in
undergraduate mathematics, with \cite{M24} introducing a library for
writing Lean proofs that resemble traditional pen-and-paper
arguments.
We refer the interested reader to \cite{BMN23}, which explores how
design choices in interaction, automation, and proof presentation can
influence students’ reasoning and their ability to structure
mathematical arguments.

These findings point to a clear conclusion: time invested in
formalising teaching material pays dividends in clarity and
pedagogical effectiveness. Formalisation yields dual rewards --
sharper theories and sharper learners—supporting the broader claim
that integrating \formalmethods into both research and teaching
enhances understanding and communication alike. Yet this mutually
beneficial relationship remains underappreciated and deserves further
exploration and attention.

%\url{https://ihp2014.pps.univ-paris-diderot.fr/doku.php?id=start}

%\newcommand{\bob}{\textsf{Bertha}}
%\newcommand{\alice}{\textsf{Alice}}

\paragraph{Technical debt}
\Formalmethods, as well as lightweight or semi-formal approaches, can serve as strategic tools in mitigating technical debt.
This potential remains underappreciated in both industrial and
academic discourse, yet it aligns closely with the foundational aims
of \formalmethods: to make critical aspects of software systems and
processes precise, analysable, and resilient to change.

Technical debt often arises when short-term development expediencies
lead to long-term maintenance challenges, typically due to ambiguous
requirements, ad hoc decision-making, loss of design rationale, or
failure to create and preserve institutional memory.
Formal specifications can directly counteract these issues by making
intended behaviours explicit and verifiable, thereby preventing
architectural drift and easing future extensions or modifications. 
These specifications act as semantic anchors that endure beyond individual codebases or team membership.
Similar considerations apply to semi-\formalmethods, albeit with laxer semantics and weaker guarantees.

Lightweight \formalmethods such as Alloy, Spec Explorer, and TLA+/TLC\footnote{TLA+ can be used in a lightweight fashion, especially when applying its bounded model checker TLC. However, its full logic and TLAPS proof assistant position fall under the heavyweight category. In short, TLA+ has range.}, are particularly well suited to this purpose. 
Their modest barrier to entry makes them ideal for early-phase modelling, where much technical debt originates.  
Partial formalisation has been shown to surface ambiguities and design flaws early in the lifecycle, significantly reducing the need for later refactorings.
As noted in \cite{JW96}, 
%\begin{quote}
\enquote{for everyday software development, the purpose of formalisation is to reduce the risk of serious specification and design errors.} 
%\end{quote}
Such early detection significantly curtails the frequency and complexity of later refactorings, thereby constraining technical debt.

Moreover, \formalmethods foster institutional memory. 
Comparable to design patterns -- which are known to facilitate shared understanding and communication in development teams \cite{KS00} -- formal specifications provide enduring artefacts that explicitly encode assumptions, invariants, and rationale. 
These artefacts prove invaluable during onboarding, audits, or system maintenance, particularly in safety-critical or compliance contexts.

Recent case studies demonstrate how specification artefacts reduce
long-term maintenance costs and serve as robust documentation even as
teams and technologies evolve.
\begin{itemize}
  \item
  At AWS, engineers found that TLA+ specifications provide \enquote{a valuable short, precise, and testable description of an algorithm} and continued to refine them from prose into executable form to support future architectural change \cite{Newcombe2015}. 
  \item 
  In the Apache ZooKeeper project, developers described TLA+ specifications as a \enquote{super-doc} that not only eliminated ambiguity but also guided deeper testing and sustained system evolution \cite{10.1007/978-981-99-8664-4_11}.
  \item
  At Microsoft, the Foundations of Software Engineering group applied Spec Explorer to model and test the components of Windows and .NET Framework and report that the use of model-based testing exposed corner-case errors early, and the resulting models acted as live documentation \cite{veanes2008}. 
  This helped maintain design integrity across updates, reducing technical debt accumulation in a large and long-lived codebase.
  \item 
  Another illustrative instance is OpenComRTOS, a real-time embedded operating system where formal modelling enabled a system $10\times$ smaller than comparable RTOSes and supported junior developers in confidently implementing and maintaining complex features. 
  As one report notes, TLA+ was used as an \enquote{architectural design tool} that made it \enquote{possible for junior developers to contribute [\dots] to the complex system} and contributed to 
  correct-by-design properties \cite{10.1007/978-1-4419-9736-4_5}.
  \item 
  Alloy has been used to assist during program and configuration refactorings by modelling changes thus helping in maintaining the integrity of the software system during critical updates or migrations, and reducing the risk of accumulating technical debt due to inconsistencies between models and code \cite{massoni2011, massoni2008}.
\end{itemize}
% Similarly, Alloy models have been used to prototype and clarify design choices -- particularly in distributed protocols and data schemas -- acting as formalised documentation.

%Yet empirical evidence linking formal artefacts to \emph{software maintainability} remains sparse. 
%Practitioner surveys report that proof results themselves were maintainable, but not necessarily reusable showing a focus on artefact quality rather than their impact on code evolution~\cite{gleirscher2020survey}. 
%This suggests a critical research gap: we must investigate whether formal specifications actually reduce code churn, refactoring needs, or fault rates in the systems they model.

\Formalmethods, then, deserve attention not only as tools for verification, but also as instruments of sustainable software engineering. 
Their potential to constrain and manage technical debt may be among their most practically valuable, yet least leveraged, capabilities.
This potential remains largely anecdotal, however, and more empirical
work is needed to quantify the effects of \formalmethods on
maintainability metrics such as code churn, refactoring effort, and
architectural degradation.

% Consider the following scenarios. %
%\todo{A: Jerky flow... is this an argument advocating for (mathematical)
%  models and different levels of abstraction?}
%\todo{Gio: Agreed. Flow is far from Tupac and Biggie.
%
%One option: leave only the story of scenario 2 to get faster to the point.}
%
%Scenario $1$: developer~\bob\ is concerned with a safety
%property~$\phi$, and writes a series of tests and runs them against a
%distributed system~$S$, finding only one code execution that
%falsifies~$\phi$.
%In~$1$ week of work \bob\ amends the code to pass the failed test,
%and conclude that~$S$ enjoys~$\phi$.
%Then $\bob$ changes leaves the company.
%
%Scenario $2$: developer~\alice\ is concerned with a liveness
%property~$\psi$.
%Recalling from \cite{10.5555/114891,DBLP:journals/cacm/CookPR11} that
%well-founded induction is a proof technique that fits the purpose,
%\alice\ pens a model of~$S$ using the~$\pi$-calculus \cite{faqPi2002}, and
%after six months of work produces a report in which these mathematical
%tools are used explain why (a model of) $S$ enjoys~$\psi$.
%Then \alice\ leaves the company.
%
%In Scenario $1$, no one left in the company knows any longer
%why~$S$ enjoys~$\phi$, where in the codebase are the hacked lines,
%and why certain tests exist at all.
%%
%In Scenario $2$, the information
%remains accessible and verifiable to any future employee with a CS
%degree, because (well-founded) induction is a proof principle thought
%in every CS syllabus that we are aware~of.

%\section{Model-Based Testing in the } 

\section{Related works}
\label{sec:related-work}
This section gathers pointers that we refer the reader to
in order to enter more deeply in any of the subject
treated in this chapter.

%%% SOFTWARE AND ECONOMY
We hope that this chapter will draw attention to the reports
by \href{https://csacademy.mst.edu/people/members/herbkrasner/}{Herb Krasner} \cite{cisq2018,cisq2020,cisq2022}.
They are a vast source of aggregated data on software
related costs, moreover they highlight the costs due to poor
software quality and they provide summaries on software failures.
A reader not satisfied with \cref{sec:costs} of this chapter,
may find more histories of recent failures in
\cite[Table 1]{cisq2018}, in \cite[Table 1]{cisq2020}, 
\cite[Section 3]{cisq2022}.

Poor quality software can expose its users to malicious actors.
\cref{sec:costs-eternalblue} showcased the global impact of a
vulnerability exploited by both criminals and state actors.
The costs related to cyber security failures are indeed object of multiple studies. 
A study by the European Commission estimated that the cost of
cybercrime to the global economy reached \money{\EUR}{5.5}{\trillion}
during 2020 \cite{eu2021cybercost}.
A recent study of the status of cybersecurity in Denmark found that
$54\%$ of danish firms experienced a serious attack in 2023 alone
\cite{pwc2023}; another study focused on defence SMEs found that even
in such a critical sector $20\%$ of the surveyed firms experienced a
serious cybersecurity incident between 2020 and 2023
\cite{stentoft2025}.
A recent study of the status of cybersecurity in the United Kingdom
estimated that the single most disruptive breach from 2024 cost on
average \GBP $1,205$ with medium and large business incurring \GBP
$10,830$ and charities \GBP $460$ \cite{cyberbreachessuvery2024}.
At the global level, IBM places the average cost of a data breach in
2024 close to \money{\USD}{5}{\million} \cite{ibm2024breachcost}.

%% RISKS
\href{https://en.wikipedia.org/wiki/Peter_G._Neumann}{Peter
  G. Neumann} is arguably the world expert on computer related risks
and their societal impact.
He pioneered discussions and public awareness of the topic via
his book \cite{DBLP:books/daglib/0079418}, via the moderation of the
Internet Forum on Risks \cite{Riskigest}, which is an immense source of
material of computer failures, and via his column ``Risks to the Public''
penned regularly since 1986
\cite{DBLP:journals/sigsoft/Neumann86c,DBLP:journals/sigsoft/Neumann24}.
\cref{sec:costs} is partly informed on \cite[Chapter 2]{DBLP:books/daglib/0079418}.

%%% FAILURES
Analysis of failures is typical in many fields.
For instance, in engineering, the Journal of \emph{Engineering Failure Analysis} is entirely dedicated to this topic,
while in economy and finance an entire library exists on the
matter \cite{libraryOfMistakes}. The closest body of knowledge to this type of research, to our knowledge, are fault taxonomies and fault repositories \cite{Hamill2009,Humbatova2020}, developed in empirical software engineering. We believe more research effort in this direction, particularly on the consequences of such faults and failures will be of significant value for the community and the society. 
For instance following the failure of the computer-aided
dispatch system of London ambulances, another implementation
of the system was deployed, which was found to  address directly
almost all the issues identified as problematic in the failure \cite{DBLP:journals/ejis/FitzgeraldR05}.

%%% DATA TO STRESS TO PRESENCE OF BUGS IN TESTS
In \cref{sec:introduction} we claimed that tests can contain bugs,
and in Exercise~(\ref{exa:flaky}) we have shown the code of a faulty test.
It is an example of {\em flaky tests}, i.e. a test that may return
different outcomes in different runs against the same software.
Flaky tests are such a wide-spread phenomenon that
\href{https://cs.gmu.edu/~winglam/}{Wing Lam} has created in $2020$
the ``International Dataset of Flaky Tests'' \cite{InternationalDatasetofFlakyTests},
and $2024$ saw the creation of a workshop dedicated to flaky tests \cite{ftw2024,ftw2025}.

%% SECTION 2
%% CHIPS AND SOFTWARE ARE INFRASTRUCTURE
In \cref{sec:infrastructure} we argued that software is infrastructure.
Similarly, \href{https://www.ri.se/en/person/johan-linaker}{J. Lin\r{a}ker} states in \cite{linaker2026} that ``digital infrastructure needs to be treated with the same seriousness as physical infrastructure such as ports, roads and power grids''; and \href{https://en.wikipedia.org/wiki/Ian_Sommerville_(software_engineer)}{I. Sommerville} in \cite[Section 1.1]{DBLP:books/lib/Sommerville9}
points out that ``We can’t run the modern world without
software. National infrastructures and utilities are controlled by
computer-based systems and most electrical products include a computer
and controlling software. Industrial manufacturing and distribution is
completely computerized, as is the financial system.''
To support these positions, in \cref{sec:infra-data} we outlined concrete
examples of software as infrastructure.
In the healthcare sector the fact that IT systems
are part of the infrastructure has been suggested for instance in
\cite{healthcare01}, and the role of IT systems in key US infrastructure
is discussed in \cite{VergeSoftwareInfrastructure1,VergeSoftwareInfrastructure2}.
In \cref{sec:infra-data} we also referred to data centres and chips as infrastructure.
The fact that global scale distributed systems are part of
modern infrastructure is stated in \cite{DBLP:journals/cacm/BasinFMNNSZZ25},
and suggested by the title of the compendium of chapters \cite{softwareinfrastructure}.
A map of the data centres and their connections in the US has been
recently published by the National Renewable Energy Laboratory
Research Hub \cite{nrel25}.
As for chips, bugs in them exist to such an extent that even
\href{https://en.wikipedia.org/wiki/Linus_Torvalds}{Linus Torvalds}
expressed %his
frustration %thus
\cite{torvalds2024}.
%% \begin{quote}
%%   I think this time we push back on the hardware people and tell them
%%   it's *THEIR* damn problem, and if they can't even be bothered to say
%%   yay-or-nay, we just sit tight.
%% \end{quote}

%%% SECTION 3 CHALLENGING CODE.

%%% ISSUES DUE TO SOFTWARE SIZES
In \cref{sec:program-analysis}, and in particular \cref{tab:software-size} we used lines of code
as metric for software size.
The problem of growing software was essentially already outlined
by E. W. Dijkstra in \cite{EWD963}:
\begin{quote}
When we had no computers, we had no programming problem either.
When we had a few computers, we had a mild programming problem.
Confronted with machines a million times as powerful, we are faced
with a gigantic programming problem.
[\ldots]
As a result, programming has become one of our most demanding intellectual activities,
requiring great clarity of expression and the utmost economy of reasoning.
This conclusion has never been refuted; it is, however, regularly
denied because of its uncomfortable implications.
\end{quote}
A discussion of metrics for software complexity can be found
in \cite[Chapter 8]{10.5555/1941689}.
The issues due to the complexity of large software systems
have been discussed in \cite{MoseleyMarks06}.
Papers that discuss how increasing software size affects various
aspects of software engineering are for instance \cite{naude2007,zhang2009distributionprogramsizesimplications}.

%% FLOATING POINTS
Code that uses floating point arithmetic is notoriously
error prone. The code snippet in Exercise~(\ref{exa:fp}) is just one
example of subtle behaviour. We refer the interested reader
to~\cite{10.1145/103162.103163} and the handbook~\cite{10.5555/1823389} for further details.
A discussion of interesting bugs is provided in Section 1.3
of that handbook.

%%% PROOFS OF CORRECTNESS
%%% HISTORICAL REASONS
The first proof of correctness of a program is probably
the one published in the paper ``Checking a large routine'' in $1949$
\cite{4640518} by that Alan Mathison Turing himself.
An overview of the history of software verification
is given in \cite{DBLP:journals/sttt/FerraraAC24}.
The content of \cref{sec:costs} substantiates Section 3.3
of that overview.

The role of formal specification of programs is discussed in
\cite[Section 12.5]{DBLP:books/lib/Sommerville9},
and the us of \formalmethods is advocated by governments \cite{whitehousepaper}
and industry \cite{Newcombe2015}.%\cite{10.1145/2699417}.

Automated static analysis, verification and \formalmethods are
discussed in \cite[Section 15.1]{DBLP:books/lib/Sommerville9}.
% THE FOLLOWING IS TRUE FOR THE 7th EDITION
%which also highlights in Section 23.1.1 that ``Static validation
%techniques are often more cost-effective than testing for discovering
%interface errors.''
A good account on the practical advantages of static
types in programming languages is \cite{empiricalPl}.
In spite of the advances of type systems, programming languages
are still far from being secure. The paper \cite{DBLP:conf/snapl/CifuentesB19}
presents an interesting analysis based on empirical data of how secure programming languages are. 

%% STATIC ANALYSIS TOOLS
We suggest the book \cite{rival:hal-02402597} as entry
point into the field of (static) program analysis.
Another excellent entry point is the book \cite{spa}.
In our experience these two books complement each other well,
for \cite{rival:hal-02402597} focuses on abstract interpretation,
that instead occupies only one chapter of \cite{spa}.
A list of existing tools for static program analysis is available at \cite{wkStaticAnalysis}.

%% FURTHER MATERIAL ON SUCCESS STORIES
TLA+ featured prominently in the success cases in industry we report in \cref{sec:advantages,sec:beyond}.
These are only a small selection and we refer the reader interested \cite{Bgli2024}.
Further examples of success stories can be found in \cite{DBLP:journals/cacm/BasinFMNNSZZ25}.
%% TODO: something about other tools we mention CompCert, Rocq/Coq, Alloy

%% FURTHER MATERIAL ON SOFTWARE ENGINEERING
\href{https://third-bit.com/}{Greg Wilson} provides an entertaining summary of empirical studies on
software development and software engineering in \cite{SEgreatestHits}.

%% SECTION 6: mindset

%% THECNICAL DEBT
\cref{sec:beyond} touched on technical debt. More material 
can be found in the proceedings of the flagship conference on the topic \cite{TechnicalDebt26}.

%% EDUCATION
The essence of deductive reasoning are proofs, and teaching mathematics
requires making students grasp what a proof is. Literature exists on
how to write mathematics in general \cite{steenrod1973write,knuth1987mathematical}
and proofs in particular \cite{lamportproofs}.
While teaching what proofs are and how to manipulate via proof assistants
falls within the scope the French national project APPAM \cite{APPAM},
the paper \cite{DBLP:journals/cacm/BasinFMNNSZZ25} advocates for the integration
of formal methods in computer science education.
%https://www.scirp.org/journal/paperinformation?paperid=51631
%\todo{Is this a good paper ?}

%% https://www.sri.com/people/peter-neumann/
%% CAN ANYBODY ACCESS THIS ARTICLE ?
%% https://www.nytimes.com/2012/10/30/science/rethinking-the-computer-at-80.html
%% ACM ENTRIES
%% https://cacm.acm.org/?s=RISKS+neumann&orderby=relevance
%% https://cacm.acm.org/opinion/qa-securing-the-risk/
%% https://cacm.acm.org/opinion/promoting-common-sense-reality-dependable-engineering/

\section{Future challenges}
\label{sec:future-work}
We gather here a series of topics that are motivated by the
data presented in this chapter, but that are out of scope with respect to
the current exposition. Note that all the topics are ultimately related to resource
allocation for software validation. Attention to this phenomenon has
already been raised in the conclusion of \cite{10.1145/3333611}.

We believe that the following topics are worth attention and answering them is a venue for future work.

% \gb{Adrian: The idea is to use the following questions about the data in Section 2
%   to lure economist into our discussion. They could provide a much deepr analysis
%   than what we did in this chapter. Can you think of one such item in the following list ?}
% How many yearly budgets of the NHS could be paid using the money
% wasted due to the CPSQ? How many more professors could be hired?

% \gb{Everybody: we should add also an item about AI. Consider the post at the URL
%   \url{https://martin.kleppmann.com/2025/12/08/ai-formal-verification.html}
%   and the very recent and impressive paper at \url{https://www.arxiv.org/abs/2601.03298}.}

\begin{description}
\item[{\bfseries\em Putting costs into perspective.}]\mbox{ }\\
  We presented the costs of poor quality software and software failures.
  To better assess these costs, these need to be compared and contrasted against
  the value created by software, 
  % (but how to evaluate it?), 
  and the costs of failures in other domains, such as civil engineering.
  
\item[{\bfseries\em The legal costs of software defects independently of catastrophic events.}]\mbox{ }\\
  For a more holistic assessment of costs incurred by software-induced failures, one needs to investigate the average cost of lawyers and attorneys paid by software houses, and identify the most common legal disputes related to software.

\item[{\bfseries\em Opportunity cost of software defects.}]\mbox{ }\\
  The resources spent debugging, patching, and firefighting could instead be devoted to developing new features, improving user experience, or pursuing strategic innovation. 
  This diversion often delays product launches, erodes customer trust, and prevents investment in activities that generate long-term competitive advantage.
  At a macro level, the cumulative cost of defective software represents a misallocation of skilled labor and capital across the economy. 
  Resources absorbed by rework, outages, and technical debt could otherwise support productivity-enhancing investments such as infrastructure, education, or research and development.

\item[{\bfseries\em The weight of mathematical proofs in court.}]\mbox{ }\\
  One needs to understand better the bearing and role of verified software and proofs in the context of a legal dispute, and determine any legislation that concerns this matter.
  % If a software is involved in a legal dispute, and the company in charge of it has {\em proven} it correct, what is the  bearing of the proof in a court?
  % Is there any legislation on this matter?

\item[{\bfseries\em The role of policies and regulatory frameworks}]\mbox{ }\\
  We need to investigate how \formalmethods play a role in policies and regulatory frameworks.
  % There are existing general frameworks e.g., regulating privacy and security (like GDPR, NIS2) and specialised regulations for critical domains (like avionics DO-178 standards) and given the potential economic, societal, and human impact of software failures, it is reasonable to expect (and we believe, request) increased attention.
  The verification community needs to identify how it contributes technical expertise to evidence-based regulation, anticipating potential risks, and co-developing governance models that strike a balance between fostering innovation and ensuring accountability. For instance, one could codify safeguards without locking into specific technical solutions or limiting their effectiveness by lack of precision.
  % Furthermore, future research agendas should incorporate mechanisms for translating technical results into insights accessible to policy stakeholders, thereby bridging the gap between scientific progress and practical governance.

\item[{\bfseries\em The definition of metrics to assess the feasibility of proof
    techniques.}
  % (analogously to computational complexity to assess %the execution time of algorithms):
]\mbox{ }\\
More work needs to be conducted to assess how techniques that work on paper and programs of few lines can scale in order to be employed by automated tools 
on real software.

\item[{\bfseries\em The economical impact of lightweight \formalmethods for software verification.}]\mbox{ }\\
  The verification community needs to understand better the effectiveness and tradeoffs of using lightweight \formalmethods as a method to mitigate the costs induced by verification on software development.
  % due to software defects?

\item[{\bfseries\em Formal Verification and AI}]\mbox{ }\\
Recent applications of Large Language Models (LLMs) to theorem proving~\cite{urban2026130klinesformaltopology} suggest that formal verification should become vastly cheaper, changing the economics of software development. 
At the same time, the proliferation of AI-generated code will need formal verification counteracts the imprecise and probabilistic nature of LLMs and ensure that it works as intended.   
\end{description}

\section{Conclusion}
\label{sec:conclusion}

The literature on software systems is vast, and it contains excellent
publications on the economical impact of software \cite{cisq2018,cisq2020,cisq2022},
the risks related to it \cite{DBLP:books/daglib/0079418},
its engineering \cite{DBLP:books/lib/Sommerville9,10.5555/1941689},
and its analysis \cite{rival:hal-02402597}.
Each publication, though, is aimed at a different part of the community,
and focuses on a single subject matter.

%%% CONTRIBUTIONS
%%% CONTRIBUTIONS
To the best of our knowledge, this chapter is the first attempt
to ``connect all the dots'': it presents an exposition that goes
from the societal importance of software (\cref{sec:infrastructure}),
to the success stories of formal verification (\cref{sec:mitigation}),
while sketching via series of examples the difficulties of
understanding seemingly trivial code snippets (\cref{sec:program-analysis}),
giving thorough attention to the the actual costs and root causes
of software failures (\cref{sec:costs}), and recalling also potential
costs of two kinds of software failures (\cref{sec:potential-costs}).

%presents the most
%detailed yet uniform analysis of recent software failures that we
%are aware of.

We hope that this chapter made the reader better grasp the importance
of formal verification and program analysis {\em for society}.
We believe this importance is ultimately the reason why
\begin{inparaenum}[(a)]
\item students (i.e. future practitioners) should invest time mastering
  the skills that underpin formal verification in particular,
  and \formalmethods in general;
\item research in \formalmethods and verification should be adequately funded
  (even the White House states that ``further innovation in approaches
  to make \formalmethods widely accessible is vital to accelerate
  broad adoption.'' \cite{whitehousepaper}).
\end{inparaenum}
The reader will judge whether we answered questions~(\ref{question1},\ref{question}).

As for teaching, our experience is that discussing software failures and their
real costs help motivating students. Even more so if the discussion is
complemented with an analysis of the top software weakness and the examples
provided by MITRE \cite{cwe2025}.

We conclude recalling from \cite{brooker2025} %\cite{10.1145/3729175}
the following points:
``The challenge is educating [\ldots] developers [\ldots], teaching
the art of rigorous thinking.''; and ``The education gap begins at the
academic level, where even basic formal reasoning approaches are
rarely taught, making it difficult for graduates from top institutions
to adopt these tools. Although \formalmethods and automated reasoning
are crucial for industry applications, they continue to be viewed as niche fields.
We anticipate that increased industry adoption of \formalmethods
and automated reasoning will attract more talent to this domain.''

We hope these remarks will convince the reader that skills in
formal verification are not a mere theoretical exercise. They
are very much needed in practice.

\bigskip

\newcommand{\hanle}{\href{https://www.informatik.tu-darmstadt.de/se/gruppenmitglieder/groupmembers_detailseite_30784.en.jsp}{Reiner Hähnle}}

\paragraph{Acknowledgements}
The first author would like to thank \href{https://akhirsch.science/}{Andrew Hirsch}, %
\hanle, and \href{https://www.cas.mcmaster.ca/~emil/}{Emil Sekerinski} for the positive feedback at ETAPS 2025; 
\href{https://fr.linkedin.com/in/filippo-biondi95/en}{Filippo Biondi} for the feedback on the MCAS;
\href{https://researchprofiles.tudublin.ie/en/persons/paul-hynds-2}{Paul Hynds} and
\href{https://www.lacl.fr/~dvaracca/}{Daniele Varacca} for the encouragement.

The third author would like to thank Marta Carlet for her valuable
feedback and for sharing her engineering expertise in the discussion of
the Engineering, Aerospace and Defence cases in Section 4.

The need for this chapter and some of the main ideas put forth in it emerged
thanks to discussions between the first author and \href{https://www.irif.fr/~dagand/}{Pierre-Evariste Dagand},
who also contributed valuable feedback.

The presentation of the material greatly benefited from the feedback by
\href{https://www.lacl.fr/~dvaracca/}{Daniele Varacca}, \hanle\, and
\href{https://www.irif.fr/~narboux/}{Julien Narboux}.
In particular, \hanle\ helped improve the overall clarity of the material,
and he provided the information on the Zune bug outlined in
Exercice~(\ref{exa:zune}); and \href{https://www.irif.fr/~narboux/}{Julien Narboux} provided valuable
references about teaching, and Exercice~(\ref{exa:abs}) in the appendix.

\href{https://www.lamport.org/}{Leslie Lamport} and
\href{https://www.linkedin.com/in/markus-kuppe-643559180}{Markus
  Kuppe} helped with useful pointers to TLA+ and its applications.

\href{https://hugo.feree.fr/}{Hugo Férée} and Mariana Milicich helped
with meticulous proof reading, and \href{https://www.irif.fr/~guatto/}{Adrien Guatto}
corrected a fishy mistake in the code of Exercice~(\ref{exa:flaky}).

The third author has been supported by an invitation of Université Paris Cité.

%% -- Bibliography ------------------------
\bibliographystyle{spbasic2} 
%% This macro typesets the archived note for online sources
%% #1 archive date
%% #2 archive url 
%\newcommand{\archived}[2]{Archived on #1 at \url{#2}}
% \newcommand{\archived}[2]{{\href{#2}{Archived on #1}}}
% \newcommand*{\archived}[2]{}

%% Single
% \bibliography{references,sites}

%% Split

% references to bibliographic sources
%% LATEX TO PRODUCE THE CORRECT .bbl FILE
%% \begin{btSect}{references}
%% \section*{Bibliography}
%% \btPrintCited
%% \end{btSect}
%% LATEX TO MAKE ARXIV COMPILE THE BIBLIOGRAPHY
\renewcommand{\refname}{Bibliography}

% references to online sources
%% LATEX TO PRODUCE THE CORRECT .bbl FILE
%\begin{btSect}{sites}
\section*{Sitography}
Wherever possible, sources are cited together with a stable link to an archived version preserved by the Internet Archive, with the date of the specific snapshot consulted. For those few sources not captured by the archive, only the original web address is provided.
%% \btPrintCited
%% \end{btSect}
%% LATEX TO MAKE ARXIV COMPILE THE BIBLIOGRAPHY
\makeatletter
\renewcommand{\refname}{\vspace*{-\baselineskip}}
\makeatother

%% -- End Bibliography --------------------

\clearpage
\appendix
\newcommand{\side}[1]{{\small #1}}

\section{Software failures with no reported cost}
\label{sec:attacks-to-hospitals}

In this appendix we gather pointers to software failures
whose costs (economical or human) are, to the best of our knowledge, not documented.
Each subsection contains a list of titles of the articles appeared in newspapers,
and each title is a link the URL of the relevant article.

\subsection{Cyber attacks to hospitals}
The following list contains the titles of the news paper articles 
on cyber attacks to English or Canadian hospitals that we are aware of.

\begin{itemize}
\item
  \href{https://www.theguardian.com/society/2017/may/14/nhs-cancer-patients-treatment-delays-cyber-attack}{NHS cancer patients hit by treatment delays after cyber-attack}\hfill\side{2017}%
 
\item
   \href{https://www.bitdefender.com/blog/hotforsecurity/cancer-treatments-cancelled-after-canadian-hospitals-hit-by-ransomware-attack/}{Cancer treatments cancelled after Canadian hospitals hit by ransomware attack}
\hfill\side{2023}%

\item
  \href{https://www.theguardian.com/society/article/2024/jun/07/london-hospitals-cancel-cancer-surgeries-after-cyber-attack}{London
    hospitals cancel cancer surgeries after cyber-attack}%
\hfill\side{2024}%

  \item
  \href{https://www.theguardian.com/society/article/2024/jun/04/cyber-attack-london-hospitals}{Services disrupted as London hospitals hit by cyber-attack}
  \hfill\side{2024}%
  
  \item
  \href{https://www.bbc.com/news/articles/c288n8rkpvno}{Critical incident over London hospitals' cyber-attack}
\hfill\side{2024}%

\item  
\href{https://www.bbc.com/news/articles/cljj1d2nz00o}{Trainees urged to help hospitals after cyber-attack}
\hfill\side{2024}

\item
  \href{https://www.bbc.com/news/articles/cxee7317kgmo}{'Russian criminals' behind hospitals cyber attack}
\hfill\side{2024}
  
\item
  \href{https://www.theguardian.com/society/article/2024/jun/05/london-nhs-hospitals-revert-to-paper-records-in-wake-of-russian-cyber-attack}{London NHS hospitals revert to paper records after cyber-attack}
\hfill\side{2024}%
  
\item
  \href{https://www.theguardian.com/society/article/2024/jun/10/nhs-appeals-for-o-type-blood-donations-after-cyber-attack-delays-transfusions}{NHS appeals for O-type blood donations after cyber-attack delays transfusions}
\hfill\side{2024}%
\end{itemize}

\subsection{Accidents in aviation}

The following list contains links to newspaper articles
on aviation accidents due to software failures,
that have caused no casualties and have no reported overall cost.

\begin{itemize} 
\item %
  \href{https://techmonitor.ai/technology/networks/rare-data-error-in-nats-air-traffic-control-system-caused-uk-flight-chaos}{Rare data error in NATS air traffic control system caused UK flight chaos}%
\hfill\side{2023}%

\item
  \href{https://www.independent.co.uk/travel/news-and-advice/air-traffic-control-failure-what-happened-b2406337.html}{What went wrong during the UK’s crippling air traffic control failure?}
  \hfill\side{2023}%

  \item
\href{https://www.nats.aero/news/nats-report-into-air-traffic-control-incident-details-root-cause-and-solution-implemented/}{NATS report into air traffic control incident details root cause and solution implemented}
\hfill\side{2023}%

%% The report confirms that safety was maintained throughout the incident and thata solution has been implemented to avoid any possible recurrence.
%% This scenario had never been encountered before, with the system having previously processed more than 15 million flight plans over the 5 years it has been in service. Steps have been taken to ensure the incident cannot be repeated.

\item \href{https://www.caa.co.uk/publication/download/21478}{Independent Review of NATS (En Route) Plc’s Flight Planning System Failure}
  \hfill\side{2024}%
%% PAGE 9, Point 2.1
%% The cause of the failure of the NERL flight plan processing system (FPRSA-R)
%% was the inability of the system software to successfully process the flight plan data
  %% for a specific flight from Los Angeles to Paris (Orly) on 28 August 2023

\item \href{https://news.sky.com/story/serious-software-glitch-meant-plane-taking-off-from-bristol-barely-cleared-the-runway-13149265}{Software glitch meant plane barely cleared the runway}
  \hfill\side{2024}%

\end{itemize}

\subsection{Accidents in railway systems}
The following list contains links to newspaper articles
on railway accidents due to software failures,
that have caused no casualties and have no reported overall cost.
We also provide a summary on the OV-chipkaart accident.

\begin{itemize}

\item \href{https://www.computerworld.com/article/1654565/sydney-trains-blames-bungled-it-upgrade-for-transport-chaos.html}{Sydney Trains blames bungled IT upgrade for transport chaos}
  \hfill\side{2018}%

\item \href{https://www.bright.nl/nieuws/1107046/ov-chipkaart-kan-ns-kaartautomaten-laten-crashen.html}{Ov-chipkaart kan NS-kaartautomaten laten crashen}
  \hfill\side{2011}
\end{itemize}

%%
%% FULL DESCRIPTION BY ADRIAN
%% DO NOT EREASE
%%
%\parcase{The OV-chipkaart ticketing system}{sec:costs-ov-ticketing}
\paragraph{The OV-chipkaart ticketing system}
In 2010, a specially malformed character or bit-error on a
(potentially non-hacked) OV-chipcard caused NS ticket vending machines
by the Dutch OV-chipkaart system to crash~\cite{OV-chipkaart:2011}.
Users seeking to purchase a paper ticket found the machines
unresponsive until these were manually rebooted.
Security experts noted that although hackers could craft such a card
intentionally,  random data corruption could trigger the same failure. 
% indicating a lack of software robustness in input handling. 

The root cause lay in insufficient input validation and error handling
within the ticket machine software, where a single corrupted or
malformed character on a chip card exposed a software flaw that lead
to system crashes.
Apart from the lack of defensive programming measures such as
sanitizing input, another limitation exposed by this vulnerability was
the lack of graceful degradation, were the entire ticketing machine
became unresponsive following a defect in one aspect of its
operations.
Formal specifications defining permissible input data formats and
error-handling invariants would have enabled the systematic modeling
of the software behavior under all possible input conditions,
including corrupt or unexpected card payloads.
Verification methods such as reachability analysis via model checking
would then ensure that invalid characters are safely rejected without
catastrophic failure (before the system is deployed).

Although no injuries resulted from the ticketing system crashes, the
implications were tangible in operational and financial terms.
Each vending machine crash disrupted customer service and incurred
costs: staff time to reboot systems, lost ticket sales  and potential
reputational damage.
While no consolidated monetary figure was published, the incident
highlighted systemic weaknesses costly to resolve over roughly $2,500$
machines nationwide.

%% \item \href{https://news.sky.com/story/remember-the-y2k-bug-microsoft-confirms-new-y2k22-issue-12507401}{Y2K22 bug}\hfill\side{2022}%

\section{Further exercices}
\label{sec:appendix-program-analysis}

%How easy is thus to perform {\em program analysis} ?
\begin{exercice}[BASH]
  \label{exa:fork-bomb}
  What does the following BASH command line do.
  What could go wrong executing it ?
\begin{minted}[frame=single]{bash}
fork(){ fork | fork & };fork
\end{minted}
\end{exercice}

\begin{exercice}[C]
  \label{exa:abs}
  We now report an example due to David Mentré. 
  In mathematics the absolute value function on real numbers, i.e. 
  $\mathit{abs}: \mathbb{R} \longrightarrow  \mathbb{R} $ is defined thus:
  $$
  \mathit{abs}(x) = 
  \begin{cases}
    x, & \mathit{if}\ x \leq 0,\\
    -x, & \mathit{if}\ x < 0.
  \end{cases}
  $$
  To simplify the discussion, let us consider this function only over integers
  representable in a computer, i.e. typically values of type \mintinline{ocaml}{int}.
  A tentative implementation is the following code,
\begin{minted}[frame=single]{c}
int abs(int x)
{
  int z;
  if (x<0) z=-x;
  else z=x;
  return z;
}
\end{minted}
Why it is {\em not} a correct implementation ?
\end{exercice}

\begin{exercice}[C++]
  The following code from the Nintendo NEX library contains at least one bug. We encourage the reader to find it.
  \begin{minted}[frame=single]{cpp}    
void StringConversion::T2Char8(const qChar *src, char *dest,
                               size_t destLength) {
    UnicodeToUtf8(src, dest, destLength);
}
  
size_t UnicodeToUtf8(const wchar_t *src, char *dest,
                     size_t destLength) {
    char *base = dest;
    char *limit = dest + destLength - 1;
    while (dest != limit) {
        wchar_t c = *src++;
        if (!c) break;
        else if (c < 0x80) {
            *dest++ = c;
        } else if (c < 0x800) {
            *dest++ = (c >> 6) | 0xC0;
            *dest++ = (c & 0x3F) | 0x80;
        } else {
            *dest++ = (c >> 12) | 0xE0;
            *dest++ = ((c >> 6) & 0x3F) | 0x80;
            *dest++ = (c & 0x3F) | 0x80;
        }
    }
    *dest = 0;
    return dest - base; }
\end{minted}
Should the reader give up, the bug and an exploit based on it are
documented on the web page \cite{wii2024}.
%\href{https://reversing.live/hacking-hundreds-of-wii-us-at-once.html}{``Hacking Hundreds of Wii U''}
\end{exercice}

\begin{exercice}[C]
  In this example we give an {\em obfuscated} and somewhat artistic code.
  In fact, it is an entry of the  $1988$ International Obfuscated C Code Contest,
  \begin{minted}[frame=single]{c}
#define _ -F<00||--F-OO--;
int F=00,OO=00;
main(){F_OO();printf("%1.3f\n",4.*-F/OO/OO);}F_OO()
{
            _-_-_-_
       _-_-_-_-_-_-_-_-_
    _-_-_-_-_-_-_-_-_-_-_-_
  _-_-_-_-_-_-_-_-_-_-_-_-_-_
 _-_-_-_-_-_-_-_-_-_-_-_-_-_-_
 _-_-_-_-_-_-_-_-_-_-_-_-_-_-_
_-_-_-_-_-_-_-_-_-_-_-_-_-_-_-_
_-_-_-_-_-_-_-_-_-_-_-_-_-_-_-_
_-_-_-_-_-_-_-_-_-_-_-_-_-_-_-_
_-_-_-_-_-_-_-_-_-_-_-_-_-_-_-_
 _-_-_-_-_-_-_-_-_-_-_-_-_-_-_
 _-_-_-_-_-_-_-_-_-_-_-_-_-_-_
  _-_-_-_-_-_-_-_-_-_-_-_-_-_
    _-_-_-_-_-_-_-_-_-_-_-_
        _-_-_-_-_-_-_-_
            _-_-_-_
}
\end{minted}
We have compiled and run it. Its execution prints on the standard output {0.250}. Why ?
\end{exercice}
%The algorithm described by the pseudo-code in \cref{fig:hyman} appeared first in \cite{DBLP:journals/cacm/Hyman66} and,
%to the best of our knowledge, more than $10$ years later.

\section{Further material on the Horizon scandal}
\label{sec:horizon-appendix}
This appendix provides further pointers to the enormous literature
on the Horizon scandal, it gives the list of software issues
discussed during the trial, and it reports two of the very technical
remarks on the software that appear in the official documents related to the trial.

An entire book on the matter has already been published
\cite{wallis2025great}, good entry points to find official
documentation on the matter are the websites
\cite{justiceforpostmasteralliance,horizoninquiry2025}. A reader
seeking more entertainment may want to start from the TV series on the
matter \cite{wkBatesPostoffice}.

Avoiding software issues is paramount. Discussing any such issue
better be done while developing software, lest it be done in a court.
The official documentation related to the Horizon Issues trial
provides an ample example of this.

While discussing the issues of the Horizon system is complicated by
the existence of three versions of the software \cite[Section F.442]{teachAppJudg6},
\begin{enumerate}[(1)]
\item Legacy Horizon, 2000 to 2010;
  
\item Horizon Online in its HNG-X form, which ran from the introduction of Horizon
  Online in 2010 to February 2017;
  
\item Horizon Online in its HNG-A form, when it was changed to run on a different
platform, namely Windows 10;
\end{enumerate}
the software issues are summarised in the {\em Appendix $2$ Summary of
  Bugs, Errors, Defects Judgment (No. 6) ``Horizon Issues''}
\cite{appendixTwoJudg6}, which presents the following list,
\begin{inparaenum}[(1)]
\item Receipts and Payments Mis-match bug
\item Callendar Square/Falkirk bug
\item Suspense Account bug
\item Dalmellington bug/Branch Outreach Issue
\item Remming In bug
\item Remming Out bug
\item Local Suspense Account issue, not the same as 3. Suspense
  Account bug
\item Recovery Issues
\item Reversals
\item Data Tree Build Failure discrepancies
\item Girobank discrepancies
\item Counter-replacement issues
\item Withdrawn stock discrepancies
\item Bureau discrepancies
\item Phantom Transactions
\item Reconciliation issues
\item Branch Customer discrepancies
\item Concurrent logins
\item Post \& Go/TA discrepancies in POLSAP
\item Recovery Failures
\item Transaction Correction Issues
\item Bugs/errors/defects introduced by previously applied PEAK fixes
\item Bureau de change
\item Wrong branch customer change displayed
\item Lyca top up
\item TPSC 250 Report
\item TPS
\item Drop and Go
\item Network Banking Bug
\end{inparaenum}

For space reasons, we do not discuss every entry in this list, and we
encourage the reader interested in the details to consult
\cite[Section E]{teachAppJudg6} and possibly
\cite[pag. 301]{batesvspostoffice}.

Here we meekly use two of the above bugs to provided empirical evidence
of three phenomena that are ever-lasting sources of problems: mutable state,
the {\tt NULL} value, concurrency.

The {\em Suspense Account Bug} is related to a mutable state \cite[paragraph 154]{teachAppJudg6}:
''The root cause of the problem was that under some specific, rare circumstances some
temporary data used in calculating the Local Suspense was not deleted when it should
have been, and so was erroneously re-used a year later.''

The {\em  Remming In bug} is related to concurrency and {\tt NULL}
value. \cite[Paragraph 190]{teachAppJudg6} states\footnote{The fonts
here are slightly changed to enhance readability.}
``when an auto remin pouch id is settled successfully, the system updates the
{\tt COUNTER\_READ\_TIMESTAMP} in {\tt LFS\_RDC\_HEADER} to a not {\tt
NULL} value for that pouch id.
The race condition for auto remin pouch delivery is handled at\\
{\tt SettlePouchDeliveryServiceSettlementProcessor.processPouch().}\\
This method checks during settlement whether the
{\tt COUNTER\_READ\_TIMESTAMP} in {\tt LFS\_RDC\_HEADER} is {\tt NULL}
or not {\tt NULL} value.
If {\tt NULL}, the pouch id is good and settlement completes successfully.
If not {\tt NULL}, the pouch id is already processed and error is thrown.
The query that gets the {\tt COUNTER\_READ\_TIMESTAMP} from {\tt
LFS\_RDC\_HEADER} is 'SettlePouchDeliveryPreCheck'.
In this query, the input parameter for pouch id is defined
incorrectly.
It is given {\tt pouchBarcode[String]}, but in dyno the pouch id is
{\tt pouchId}.
This is the root cause why the query always returns null although the
{\tt COUNTER\_READ\_TIMESTAMP} is not {\tt NULL}.''

As a closing note, we recall a remark on one of the key players in the
Horizon scandal: International Computers Limited.
\cite[Appendix B, pag. 287]{originasofadisaster2022} describes the
attitude of the company thus:
\begin{quote}
\enquote{ICL’s whole attitude to delivery of the project has been consistently
  to find reasons not to do what is asked, but to cut corners: this
  does not bode well for a reassuring view of the future \ldots}
\end{quote}

\section{Glossary of Terms}
\label{sec:glossary}
In this section, we provide in alphabetical order a glossary %of terms
defining some of the key terms we have used in this chapter.
The definitions are mostly based on / inspired by standard definition,
e.g., from IEEE Standard 159342 \cite{IEEE159342}, IEEE CS SwEBoK V3.0 \cite{GuideToSwebok},
SEBoK V2.1.2 \cite{SEBok} and %well-known reviews \cite{WoodcockLBF09} of the field.  
reviews of the field such as \cite{WoodcockLBF09}.

\begin{figure}[t]
  \caption{A Venn diagram describing the relations between software
  validation, formal methods, and formal verification.}
\begin{center}
  \includegraphics[width=180pt]{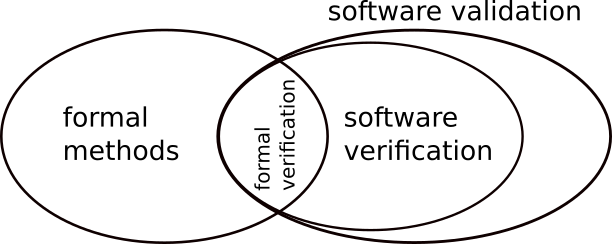}
\end{center}
\label{fig:terminology}
\hrulefill
\end{figure}

The relations between software validation, software verification and
formal verification are shown pictorially in \cref{fig:terminology}.

\paragraph{Failure} %
An undesired effect observed in the computer system’s delivered service.

\paragraph{Fault} %
The cause of a malfunction in a computer system, typically arising from an incorrect or forgotten implementation of piece of requirement.

\paragraph{Formal methods} %
A set of mathematically-founded techniques for formal specification and verification of  computer-system requirements. 

\paragraph{Formal semantics} %
A mapping from the objects (well-formed sentences) in a formally
defined syntax (see formal syntax below) into a set of mathematically
defined objects (called the semantic domain).

\paragraph{Formal specification} %
A specification written in a formal notation, i.e., a language with a
formal syntax and semantics, often for use in formal verification.

\paragraph{Formal syntax} %
The definition of well-formed sentences in a language, usually defined
using a form of grammar or (extended) regular expressions, such as the
Backus-Naur Form (BNF).

\paragraph{Formal verification} %
Using mathematical techniques to prove or disprove the properties of a computer system or its formal model. Both the properties and the model of the system are likely to be formal specifications or at least models with a formal semantics. 
%Verification by means of formal methods.
%% A specification written in a formal notation, i.e., a language with a
%% formal syntax and semantics, often for use in formal verification. 

\paragraph{Quality} %
The degree to which a computer system, component, or process meets
specified requirements and the users' expectations.

\paragraph{Quality assurance} %
A planned and systematic pattern of all actions necessary
to provide adequate confidence that an item or product conforms to an
established level of quality (see above for the definition of
quality).

\paragraph{Validation} %
Processes and techniques for evaluating whether a computer system or
component satisfies user expectations. Validation is an attempt to
ensure that the right product is built —that is, the product fulfils
its specific intended purpose.

\paragraph{Verification} %
Processes and techniques for evaluating whether a system or component satisfies a set of requirements (typically specified in another artefact at the beginning of the concerned development phase). 
 Verification is an attempt to ensure that the product is built correctly, in the sense that the output products of an activity meet the specifications imposed on them in previous activities.

\end{document}